\documentclass[12pt]{article}
\usepackage{titling}
\usepackage[utf8]{inputenc}
\usepackage{amsmath}
\usepackage{amsthm}
\usepackage{mathtools}
\usepackage{amssymb}
\usepackage{mathrsfs}
\usepackage[english]{babel}
\usepackage{changepage}
\usepackage{graphicx}
\usepackage{mathrsfs}
\usepackage{relsize}
\usepackage{cite}
\usepackage{authblk}
\usepackage{appendix}
\usepackage{longtable}
\graphicspath{ {./images/} }
\usepackage{fullpage}

\usepackage[normalem]{ulem}
\usepackage[dvipsnames]{xcolor}

\begin{document}

\title{Evolutionary game dynamics with environmental feedback in a network with two communities}
\author{\normalsize Katherine Betz$^1$,\; Feng Fu$^{2,3}$, \;and\;Naoki Masuda$^{1,4,5}$}
\date{%
    \footnotesize$^1$ Department of Mathematics, State University of New York at Buffalo, NY 14260-2900, USA\\[1ex]
\footnotesize$^2$ Department of Mathematics, Dartmouth College, Hanover, NH 03755, USA\\[1ex]
\footnotesize$^3$ Department of Biomedical Data Science, Geisel School of Medicine at Dartmouth, Lebanon, NH 03755, USA\\[1ex]
\footnotesize$^4$ Institute for Artificial Intelligence and Data Science, State University of New York at Buffalo, NY 14260-2900, USA\\[1ex]
\footnotesize$^5$ Center for Computational Social Science, Kobe University, Kobe 657-8501, Japan
}
\maketitle

\begin{abstract}
Recent developments of eco-evolutionary models have shown that evolving feedbacks between behavioral strategies and the environment of game interactions, leading to changes in the underlying payoff matrix, can impact the underlying population dynamics in various manners. We propose and analyze an eco-evolutionary game dynamics model on a network with two communities such that players interact with other players in the same community and those in the opposite community at different rates. In our model, we consider two-person matrix games with pairwise interactions occurring on individual edges and assume that the environmental state depends on edges rather than on nodes or being globally shared in the population. We analytically determine the equilibria and their stability under a symmetric population structure assumption, and we also numerically study the replicator dynamics of the general model. The model shows rich dynamical behavior, such as multiple transcritical bifurcations, multistability, and anti-synchronous oscillations. Our work offers insights into understanding how the presence of community structure impacts the eco-evolutionary dynamics within and between niches. 
\end{abstract}

\textbf{keywords}: evolutionary game theory, feedback-evolving games, oscillatory dynamics, bifurcation analysis

\section{Introduction}

Evolutionary game theory is the study of population changes driven by competition among different strategies. A recent adjustment of evolutionary game models with the aim of better representing the natural world is the inclusion of strategy-dependent feedback, specifically, environmental feedback \cite{Weitz16}. This type of game is called an eco-evolutionary game. This type of strategy-dependent feedback can be seen in many complex systems, such as ecological metacommunities \cite{Leibold18}, collectives of insect individuals \cite{Hanski11,Stella22i}, microbial populations \cite{West06, Sanchez13, Estrela18}, and human social and reproductive structures \cite{Rand17, Mullon18}.  A major question with models of eco-evolutionary game dynamics is conditions under which cooperation in a population can thrive when the payoff matrix, which we regard as the environment, is influenced by the action of players. Extensions of the original eco-evolutionary game dynamics models include the addition of finite carrying capacity \cite{Bairagya21}, renewable and decaying resources \cite{Tilman20, Wang20, Yan21}, imitation and aspiration dynamics \cite{Arefin21}, mutation of players \cite{Gong22}, reciprocity dynamics \cite{Ma24}, and extension to public goods games \cite{Shai19, Wang20, Jiang23, Han24}. The models can also be extended in terms of additional types of dynamic feedback, such as non-constant enhancement or degradation rates of the environmental variable, which depends on the payoff of players \cite{Cao21}, and global and local environment fluctuations \cite{Jiang23}.

Given that players of the game are embedded in structured populations in reality, evolutionary game models have been extended to the case of various networks \cite{Nowak, Szabo07, Perc13, Wang24}. Similarly, players involved in an eco-evolutionary game may be better interpreted to inhabit on nodes of a network. Therefore, eco-evolutionary games have been extended to the case of networks. For example, in eco-evolutionary games on regular graphs, it was found that a higher degree of the node creates oscillatory behavior in the population and that a lower degree promotes spread of cooperation \cite{Stella22, Zhang23}. Spatial networks are also commonly used for exploring how environmental feedback promotes cooperation \cite{Jin18, Szolnoki17, Wu18, Hauert19, Lin19, Wu19, Yang21, Ding23, He23, Lu21, Zhu23}.  Lastly, through the use of bimatrix payoffs, which are equivalent to the complete bipartite graph as population structure in the case of symmetric payoff matrices, periodic orbits in the state space have been proven to exist \cite{Gong18, Kawano19, Liu20, Fu22}. 

However, there are some vital gaps missing in the prior research on eco-evolutionary games on networks.  First, in complete bipartite graphs \cite{Gong18, Kawano19, Liu20,Fu22}, the players do not interact within each community. This assumption is suitable for modeling situations in which the population of players is divided into two different roles but otherwise not in general. Second, in most of the previous studies, the environmental state is assumed to be either a globally shared variable \cite{Gong18, Kawano19, Liu20,Fu22, Stella22, Zhang23} or local to each node (i.e., player) \cite{Jin18, Szolnoki17, Wu18, Hauert19, Lin19, Wu19, Yang21, Ding23, He23, Lu21, Zhu23}. However, it may be more realistic to assume that the environment is shared across some, but not all, players \cite{Guimaraes20, Fahimipour22}. For example, a meta-community in ecological systems may be an appropriately sized unit for considering an environmental variable \cite{Holyoak09, Brechtel18, Leibold18, Gross20}. Other eco-evolutionary game models assume network structure and assign a local environmental variable to each edge between a pair of players \cite{He23, Zhu23}.

In the present study, we extend a previously proposed model of eco-evolutionary dynamics \cite{Weitz16} to the case of networks with equally sized two communities.  Unlike the complete bipartite graph models proposed in \cite{Gong18, Kawano19, Fu22} where the players in each community only interact with those in the other community, we assume that players not only interact with those in the other community but also with those in the same community.  Next, we assume that the state of the environment depends on the type of edge in the network, similarly to \cite{He23, Zhu23}. We crucially assign one environmental variable to each type of edge, i.e., the edges within the first community, those within the second community, and those connecting the two communities. In this manner, we model the situation in which two players forming an edge may improve or deteriorate their shared environment, which is assumed to be on the edge. We do not distinguish between edges of the same type because of the symmetric population structure assumed. Unlike the previous studies similarly assuming edge-dependent environmental states \cite{He23, Zhu23}, our two-community network model, which is a minimal network model, allows analytical investigations.

Our paper is organized as follows. In Section~\ref{sec:model}, we describe our model in detail and focus on eco-evolutionary dynamics with two network communities. In Sections~\ref{sec:3dim1} and \ref{sec:3dim2}, respectively, we present our stability analysis of the simplified replicator dynamics resulting from different symmetry assumptions. In Section~\ref{sec:5dim}, we numerically investigate the rich dynamical behavior of the general model. Finally, we discuss contributions of the current work along with an outlook for future work.

\section{Model}\label{sec:model}

\begin{figure} 
\centering 
\includegraphics[width=90mm]{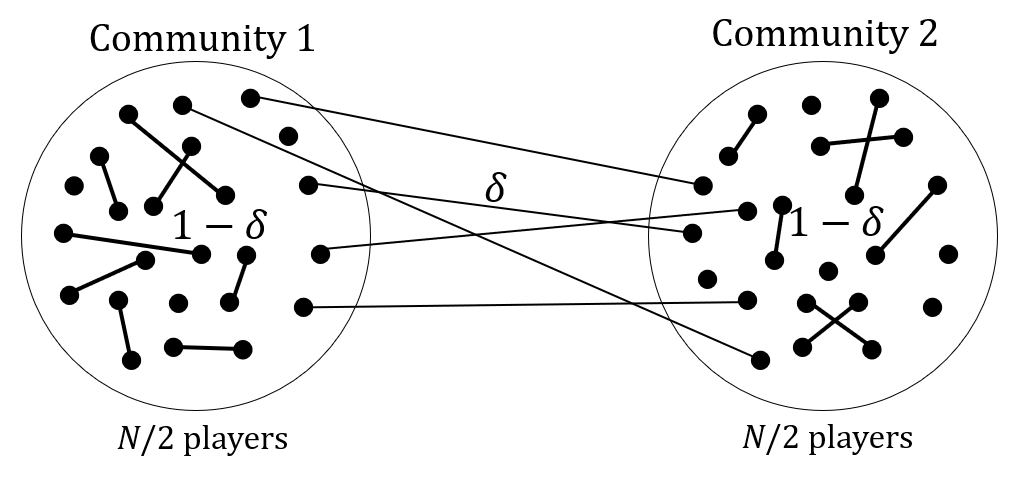} 
\caption{Schematic of the two-community network. A filled circle represents a player. Two players from the same community interact at rate $1-\delta$. Two players from the opposite communities interact at rate $\delta$. Without loss of generality, we normalize the rate parameter $0< \delta < 1$. We only show some edges for visualization purposes.}
\label{fig:twocomfig}
\end{figure}

Consider an eco-evolutionary game in a population composed of two communities.  Each player chooses either of the two actions, i.e.,  cooperation or defection. We assume that there are $N$ players in total and $N/2$ players in each community. We assume that the entire population is infinite (i.e., $N\to\infty$) and that the players interact with each other player within the same community at rate $1-\delta$ and with each player in the other community at rate $\delta>0$. See Fig.~\ref{fig:twocomfig} for a schematic. 

We consider replicator dynamics for a population on the two-community network with feedback-evolving games.  Crucially, we assume that the state of the environment depends on the type of edge in the network. We denote by $n_1 \in [0, 1]$ the state of the environment in community $1$, representing the edges within community $1$, by $n_2 \in [0, 1]$ the state of the environment in community $2$, and by $n_{12} \in [0, 1]$ the state of the environment used when a player in community $1$ and one in community $2$ interact. The environment-dependent payoff matrices for community $1$, $2$, and in between are assumed to be given by
\begin{equation}
A(n) = (1-n)\begin{pmatrix}R_0&S_0\\ T_0&P_0\end{pmatrix}+n\begin{pmatrix}R_1&S_1\\ T_1&P_1\end{pmatrix}\label{1}, 
\end{equation}
where $n$ is either $n_1$, $n_2$, or $n_{12}$.  
We assume that, if $n=0$, then cooperation is the unique Nash equilibrium, i.e., $R_0>T_0$ and $S_0>P_0$.  If $n=1$, then defection is the unique Nash equilibrium, i.e., $R_1<T_1$ and $S_1<P_1$. We label the prior inequalities as 
\begin{align}
R_0>T_0,\;S_0>P_0,\;&\;R_1<T_1,\;S_1<P_1.\label{inequal}
\end{align}

Let us define $\textbf{q}_1$ and $\textbf{q}_2$ as the two-dimensional payoff vector for a player in community $1$ and $2$, respectively. The first entry of the vector is the payoff for a cooperator. The second entry of the vector is the payoff for a defector. Define $x$ and $y$ as the fraction of cooperators in community $1$ and $2$, respectively. The fraction of defectors in community $1$ and $2$ is $1-x$ and $1-y$, respectively. We obtain
\begin{align}
\textbf{q}_1&=(1-\delta)A(n_1)\textbf{x}+\delta A(n_{12})\textbf{y},\label{q1}\\
\textbf{q}_2&=(1-\delta)A(n_2)\textbf{y}+\delta A(n_{12})\textbf{x},\label{q2}
\end{align}
where $\textbf{x} = \begin{pmatrix}x&1-x \end{pmatrix}^{\top}$,
$\textbf{y} =\begin{pmatrix}y&1-y \end{pmatrix}^{\top}$, and
$^\top$ denotes the transposition.
The first term on the right-hand side of Eqs.~\eqref{q1} and \eqref{q2} is the payoff obtained by playing with the other players in the same community. The second term is the payoff obtained by playing with the players in the opposite community.

We assume that the competition between cooperation and defection occurs only within each community because players inhabiting different communities may perceive the different environments due to the different state of the environment. Then, the replicator dynamics are given by
\begin{align}
\dot{x} &=x(1-x)(q_{11} - q_{12}),\label{x1}\\
\dot{y} &=y(1-y)(q_{21} - q_{22}),\label{y1}
\end{align}
where $q_{11}$, $q_{12}$, $q_{21}$, and $q_{22}$ are defined by 
\begin{equation}\label{qeq}
\textbf{q}_i = \begin{pmatrix} q_{i1} \\ q_{i2} \end{pmatrix}
\end{equation}
with $i \in \{1, 2 \}$. 

We give the dynamics of the environmental state of each type of edge by
\begin{align}
\dot{n}_1&=n_1(1-n_1) \left[\theta_1x-(1-x)\right], 
\label{n1}\\
\dot{n}_2&=n_2(1-n_2) \left[\theta_2y-(1-y)\right], 
\label{n2}\\
\dot{n}_{12}&=n_{12}(1-n_{12}) \left[ \theta_{12}z-(1-z) \right], 
\label{n12}
\end{align}
where $z$ is the fraction of cooperators in the entire population, i.e., $z \equiv (x+y)/2$, and $\theta_1>0$, $\theta_2>0$, and $\theta_{12}>0$ are the ratio of enhancement to degradation of the environmental variable for the respective edge type. For example, if $\theta_1$ is large, then enhancement of the environment in community $1$ occurs at a relatively small fraction of cooperators, $x$.

We let $R_3=R_0-R_1$, $T_3=T_0-T_1$, $P_3=P_0-P_1$, and $S_3=S_0-S_1$. Then, we obtain the five-dimensional dynamical system given by
\begin{align}
\dot{x} =& x(1-x)[(R_0-T_0-S_0+P_0)x+S_0-P_0 \nonumber\\
	 & -n_1(S_3-P_3)-\delta[(R_0-T_0-S_0+P_0)(x-y) \nonumber\\
	 & -(R_3-T_3-S_3+P_3)(n_1x-n_{12}y)-(S_3-P_3)(n_1-n_{12})]], \label{5x}\\
\dot{y} =& y(1-y)[(R_0-T_0-S_0+P_0)y+S_0-P_0 \nonumber\\
	 &-n_2(S_3-P_3)-\delta[(R_0-T_0-S_0+P_0)(y-x) \nonumber\\
	 & -(R_3-T_3-S_3+P_3)(n_2y-n_{12}x)-(S_3-P_3)(n_2-n_{12})]], \label{5y}
\end{align}
and Eqs.~\eqref{n1}, \eqref{n2}, and \eqref{n12}. 	 

\section{Three-dimensional system with $\theta_1\neq \theta_{12}$} \label{sec:3dim1}

In this section, we assume that $\theta_1=\theta_2$, and that the initial condition satisfies $x=y$ and $n_1=n_2$. Then, $x=y$ and $n_1=n_2$ hold true for any $t>0$. We further assume that $\theta_1=\theta_2\neq\theta_{12}$. In this case, the original five-dimensional dynamical system is reduced to the three-dimensional dynamical system given by
\begin{align}
\dot{x}&=x(1-x)[(R_0-T_0-S_0+P_0)x+S_0-P_0 \nonumber\\
 &\;\;\;-\left[n_1(1-\delta)+\delta n_{12}\right][(R_3-S_3-T_3+P_3)x+(S_3-P_3)]],\label{3x}\\
\dot{n}_1&=n_1(1-n_1)[\theta_1x-(1-x)],\label{3n1}\\
\dot{n}_{12}&=n_{12}(1-n_{12})[\theta_{12}x-(1-x)]\label{3n12}.
\end{align}
We analyze the equilibria and dynamics of this three-dimensional dynamical system. The Jacobian of this dynamical system is given by
\begin{equation}
\hspace*{-1.5cm}
\label{3Jac}
J(x,n_1,n_{12})=\begin{pmatrix}
x(1-x)\frac{\partial g}{\partial x} & x(1-x)\frac{\partial g}{\partial n_1} & x(1-x)\frac{\partial g}{\partial n_{12}}\\
+(1-2x)g(x,n_1,n_{12})&&\\
 & -n_1(\theta_1x+x-1) & \\
n_1(1-n_1)(1+\theta_1)&+(1-n_1)(\theta_1x+x-1)&0\\
&&-n_{12}(\theta_1x+x-1)\\
n_{12}(1-n_{12})(1+\theta_1)&0&+(1-n_{12})(\theta_1x+x-1)
\end{pmatrix},
\end{equation}
where
\begin{equation} \label{geq}
g(x,n_1,n_{12})=q_{11}-q_{12},
\end{equation}
and $q_{11}$ and $q_{12}$ are given by Eq.~\eqref{qeq}.

\subsection{Corner equilibria}\label{sec:C1}

We denote by $x^\ast$ the equilibrium of $x$ and similar for the other dynamical variables.
By setting $x^\ast$, $n_1^\ast$, and $n_{12}^\ast$ to $0$ or $1$, specifying the corners of the unit cube defined by $0\le x, n_1, n_{12} \le 1$,
we obtain $8$ corner equilibria. We show in Appendix \ref{app:C1} that each corner equilibrium is a saddle.

\subsection{Interior equilibria}\label{sec:I1}

In this section, we seek interior equilibria, i.e., those in which $0<x^\ast$, $n_1^\ast$, $n_{12}^\ast<1$.  By setting $\dot{n}_1 = 0$ and $\dot{n}_{12} = 0$ in Eqs.~\eqref{3n1} and \eqref{3n12}, respectively, we obtain $x^\ast=\frac{1}{1+\theta_1}$ and $x^\ast=\frac{1}{1+\theta_{12}}$, which is a contradiction, because we assumed $\theta_1 \neq \theta_{12}$.  Therefore, there are no internal equilibria.

\subsection{Edge equilibria}\label{sec:E1}

Let us examine possible equilibria on the edge of the unit cube, which we call edge equilibria. At an edge equilibrium, one variable out of $x^\ast$, $n_1^\ast$, or $n_{12}^\ast$ is between $0$ and $1$, and the other two variables are either $0$ or $1$. If $x^\ast = 0$ or $1$, then Eqs.~\eqref{3n1} and \eqref{3n12} imply that
$n_1^\ast, n_{12}^\ast \in \{0, 1\}$, leading to corner equilibria. Therefore, there is no edge equilibrium satisfying $x^\ast \in \{ 0, 1 \}$. Therefore, we search for edge equilibria such that $0<x^\ast<1$ and $n_1, n_{12}\in\{0,1\}$. Pairs $(n_1, n_{12}) = (0, 0)$ and $(1, 1)$ violate Eq.~\eqref{inequal}. The other two pairs, i.e., $(n_1, n_{12}) = (0, 1)$ and $(1, 0)$, provide equilibria.

The edge equilibrium $(x^\ast, n_1^\ast, n_{12}^\ast)=\left(\frac{P_0-S_0-\delta(P_0-P_1-S_0+S_1)}{R_0-T_0-S_0+P_0-\delta\gamma},0,1\right)$, where
\begin{equation}
\gamma = R_0-R_1-T_0+T_1-S_0+S_1+P_0-P_1,\label{gamgam}
\end{equation}
is stable if and only if 
\begin{align} 
(P_0 - S_0) (R_1 - T_1)&>(P_1 - S_1) (R_0 - T_0),\label{stabineq} \\ 
\delta_{\rm c,1} < &\delta < \delta_{\rm c,2},\label{delline1}
\intertext{and}
\theta_{12}&>\theta_1,\label{t12gtt1}
\end{align}
where
\begin{align}
\delta_{\rm c,1} \equiv& \frac{R_0-T_0-P_0\theta_1+S_0\theta_1}{\rho_1},\\
\delta_{\rm c,2} \equiv& \frac{R_0-T_0-P_0\theta_{12}+S_0\theta_{12}}{\rho_{12}},\\
\rho_1=& R_0-R_1-T_0+T_1-P_0\theta_1+P_1\theta_1+S_0\theta_1-S_1\theta_1\label{r1},\\
\rho_{12}=& R_0-R_1-T_0+T_1-P_0\theta_{12}+P_1\theta_{12}+S_0\theta_{12}-S_1\theta_{12}.\label{r12}
\end{align}
If either Eq.~\eqref{stabineq}, \eqref{delline1}, or \eqref{t12gtt1} is not met, the equilibrium is unstable. We derive Eqs.~\eqref{stabineq}, \eqref{delline1}, and \eqref{t12gtt1} in Appendix \ref{app:E1}.

The equilibrium $(x^\ast, n_1^\ast, n_{12}^\ast)=\left(\frac{P_1-S_1+\delta(P_0-P_1-S_0+S_1)}{R_1-T_1-S_1+P_1+\delta\gamma},1,0\right)$ is stable if and only if Eq. \eqref{stabineq}, 
\begin{equation} 
\delta_{\rm c,3} < \delta < \delta_{\rm c,4},\label{delline2}
\end{equation}
and
\begin{equation}
\theta_{12} < \theta_1,\label{t12ltt1}
\end{equation}
where
\begin{equation}
\delta_{\rm c,3} \equiv \frac{-R_1+T_1+P_1\theta_1-S_1\theta_1}{\rho_1}
\end{equation}
and
\begin{equation}
\delta_{\rm c, 4} \equiv \frac{-R_1+T_1+P_1\theta_{12}-S_1\theta_{12}}{\rho_{12}},
\end{equation}
hold true. If either Eq.~\eqref{stabineq}, \eqref{delline2}, or \eqref{t12ltt1} is not met, the equilibrium is unstable.  The derivation is given in Appendix \ref{app:E1}.
  
\subsection{Face equilibria}\label{sec:F1}

In this section, we seek equilibria on the face of the unit cube, i.e., those in which just one of $x^\ast$, $n_1^\ast$, or $n_{12}^\ast$ is either $0$ or $1$ and the other two are between $0$ and $1$. We call these equilibria face equilibria. Similarly to the case of the edge equilibria, if we let $x^\ast=0$ or $1$, then we obtain a corner equilibrium.  Therefore, we assume that $0<x^\ast<1$. By setting just one of $n_1^\ast$ or $n_{12}^\ast$ to $0$ or $1$, we obtain the four face equilibria shown in Table~\ref{3dimface1}. 

\begin{table}
\caption{Face equilibria of the three-dimensional dynamics when $\theta_1 \neq \theta_{12}$. We recall that $\rho_1$ and $\rho_{12}$ are defined by Eqs. \eqref{r1} and \eqref{r12}, respectively.}
\center
\def\arraystretch{1.5}
\begin{tabular}{|c |c |c|} 
 \hline
 $x^\ast$ &  $n_1^\ast$ &$n_{12}^\ast$ \\
 \hline\hline
 $\frac{1}{1+\theta_1}$ &$\frac{R_0-T_0-P_0\theta_1+S_0\theta_1}{(1-\delta)\rho_1}$  & 0  \\ 
 \hline
$\frac{1}{1+\theta_{12}}$ & 1 & $\frac{R_1 - T_1 - P_1\theta_{12} + S_1\theta_{12} +\delta\rho_{12}}{\delta \rho_{12}}$\\  
\hline
$\frac{1}{1+\theta_1}$ & $\frac{R_0 -T_0- P_0\theta_1 + S_0\theta_1 -\delta\rho_1}{(1- \delta) \rho_1}$ & 1 \\ 
 \hline
 $\frac{1}{1+\theta_{12}}$ &0  &$\frac{R_0-T_0-P_0\theta_{12}+S_0\theta_{12}}{\delta\rho_{12}}$ \\ 
 \hline
\end{tabular}
\label{3dimface1}
\end{table}


For the equilibrium $(x^\ast, n_1^\ast, n_{12}^\ast)=\left(\frac{1}{1+\theta_1},\frac{R_0-T_0-P_0\theta_1+S_0\theta_1}{(1-\delta)\rho_1},0\right)$, the Jacobian is given by
\begin{equation} \label{3J120}
J =
\begin{pmatrix}
J^{(1)}_{11} &J^{(1)}_{12}  & J^{(1)}_{13}\\
J^{(1)}_{21}&0&0\\
0&0&J^{(1)}_{33}
\end{pmatrix},
\end{equation}
where
\begin{align}
\hspace*{-1cm}
J^{(1)}_{11}&=\frac{\left[(P_1 - S_1) (R_0 - T_0)-(P_0 - S_0) (R_1 - T_1)\right]\theta_1}{(1 + \theta_1) \rho_1},\label{53}\\
J^{(1)}_{12}&=\frac{-(1-\delta) \theta_1 \rho_1}{(1 + \theta_1)^3}\label{55},\\
J^{(1)}_{13}&=\frac{-\delta \theta_1 \rho_1}{(1 + \theta_1)^3},\label{56}\\
J^{(1)}_{21}&=\frac{-(1 + \theta_1) \left[R_0 - T_0 - (P_0 - S_0) \theta_1\right] (R_1- T_1- P_1 \theta_1 +S_1 \theta_1+\delta\rho_1)}{(1- \delta)^2 \rho_1^2}\label{54},\\
J^{(1)}_{33}&=\frac{\theta_{12}-\theta_1}{1+\theta_1}\label{57}.
\end{align}
The characteristic equation is given by,
\begin{equation}
\det(J - \lambda I)=\left(J^{(1)}_{33}-\lambda\right)\left(\lambda^2-J^{(1)}_{11}\lambda-J^{(1)}_{12}J^{(1)}_{21}\right)=0\label{3char2}.
\end{equation}
Eigenvalue $\lambda_1 = J^{(1)}_{33} = \frac{\theta_{12}-\theta_1}{1+\theta_1}$ is negative if and only if $\theta_{12}<\theta_1$ (i.e., Eq.~\eqref{t12ltt1}).
The other two eigenvalues, denoted by $\lambda_2$ and $\lambda_3$,
are solutions of $\lambda^2-J^{(1)}_{11}\lambda-J^{(1)}_{12}J^{(1)}_{21} = 0$.
%
The real part of $\lambda_2$ and $\lambda_3$ is negative if and only if $-J^{(1)}_{11}>0$ and $-J^{(1)}_{12}J^{(1)}_{21}>0$. Equation~\eqref{inequal} guarantees that both $\rho_1$ and $\rho_{12}$ are positive. Therefore, $-J^{(1)}_{11}>0$ if and only if Eq. \eqref{stabineq} holds true.
Equation~\eqref{55} combined with $\rho_1 > 0$ implies that $J^{(1)}_{12} < 0$. Therefore, $-J^{(1)}_{12}J^{(1)}_{21}>0$ if and only if $J^{(1)}_{21} > 0$, which holds true if and only if
\begin{equation}
\label{deltan120}
\delta < \delta_{\rm c,3}.
\end{equation}
Note that, in Eq.~\eqref{54}, $R_0 - T_0 - (P_0 - S_0) \theta_1 > 0$ because $R_0-T_0>0$ and $P_0-S_0<0$. Therefore, this equilibrium is stable if and only if Eqs. \eqref{stabineq}, \eqref{t12ltt1}, and \eqref{deltan120} hold true.

\begin{figure}[!t]
\centering 
\includegraphics[width=130mm]{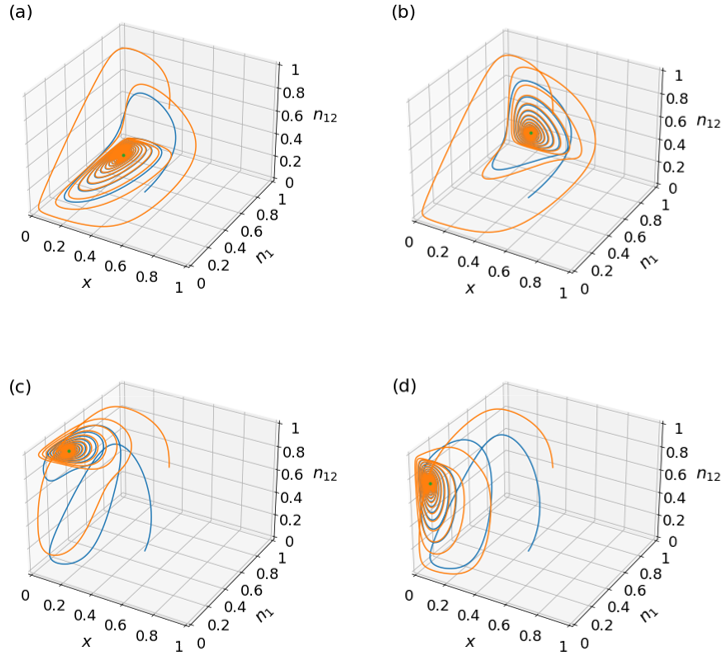}  
\caption{Convergence to face equilibria. Shown are numerically obtained trajectories of the three-dimensional system given by Eqs. \eqref{3x}, \eqref{3n1}, and \eqref{3n12}.  The green dots represent the face equilibria given in Table \ref{3dimface1}. We use the payoff matrices given by Eq.~\eqref{nummatri} and initial conditions $(x,n_1,n_{12})=(0.5, 0.4, 0.1)$
and $(0.6, 0.5, 0.8)$, of which the corresponding trajectories are shown in blue and orange, respectively. (a) $\theta_1=8$, $\theta_{12}=5$, and $\delta=0.6$. (b) $\theta_1=8$, $\theta_{12}=5$, and $\delta=0.8$. (c) $\theta_1=5$, $\theta_{12}=8$, and $\delta=0.2$. (d) $\theta_1=5$, $\theta_{12}=8$, and $\delta=0.4$.
}
\label{fig:traj}
\end{figure}

For numerical demonstration, we set
\begin{equation}\label{nummatri}
\begin{pmatrix}R_0&S_0\\T_0&P_0\\ \end{pmatrix}=\begin{pmatrix}5&1\\3&0\\ \end{pmatrix}\;\;\text{and}\;\;\begin{pmatrix}R_1&S_1\\T_1&P_1\\ \end{pmatrix}=\begin{pmatrix}3&0\\8&2\\ \end{pmatrix},
\end{equation}
which satisfy Eq. \eqref{inequal}. We also set $\theta_1=8$, $\theta_{12}=5$, and $\delta=0.6$, yielding $\lambda_1=-\frac{1}{3}$ and $\lambda_{2, 3} = -0.014 \pm 0.437i$. We show two numerically simulated trajectories starting from different initial conditions in Fig.~\ref{fig:traj}(a). Figure \ref{fig:traj}(a) indicates that the trajectories spiral into the presently discussed face equilibrium.  

The derivation of the conditions for stability of the other three face equilibria is similar; see Appendix~\ref{app:F1} for the derivation.

Equilibrium $(x^\ast, n_1^\ast, n_{12}^\ast)=\left(\frac{1}{1+\theta_{12}},1,\frac{R_1 - T_1 - P_1\theta_{12} + S_1\theta_{12} +\delta\rho_{12}}{\delta \rho_{12}}\right)$ is stable if and only if
Eqs. \eqref{stabineq}, \eqref{t12ltt1}, and
\begin{equation}\label{deltan11}
\delta > \delta_{\rm c,4}
\end{equation}
hold true. For numerical demonstration of this face equilibrium, we set $\theta_1=8$, $\theta_{12}=5$, and $\delta=0.8$, yielding $\lambda_1=-\frac{1}{2}$ and $\lambda_{2, 3} = -0.019 \pm 0.232i$. As expected, Fig.~\ref{fig:traj}(b) shows that two trajectories starting from different initial conditions spiral into the presently discussed face equilibrium.  

Equilibrium $(x^\ast, n_1^\ast, n_{12}^\ast)=\left(\frac{1}{1+\theta_1},\frac{R_0 -T_0- P_0\theta_1 + S_0\theta_1 -\delta\rho_1}{(1- \delta) \rho_1}, 1\right)$ is stable if and only if
\eqref{stabineq}, \eqref{t12gtt1}, and
\begin{equation}\label{deltan121}
\delta < \delta_{\rm c,1}
\end{equation}
hold true. For numerical simulations, we set $\theta_1=5$, $\theta_{12}=8$, and $\delta=0.2$, yielding $\lambda_1=-\frac{1}{2}$ and $\lambda_{2, 3} = -0.019 \pm 0.554i$. As expected, two trajectories, shown in Fig.~\ref{fig:traj}(c), spiral into the presently discussed face equilibrium. 

Equilibrium $(x^\ast, n_1^\ast, n_{12}^\ast)=\left(\frac{1}{1+\theta_{12}},0,\frac{R_0-T_0-P_0\theta_{12}+S_0\theta_{12}}{\delta\rho_{12}}\right)$ is stable if and only if
Eqs.~\eqref{stabineq}, \eqref{t12gtt1}, and
\begin{equation}\label{deltan10}
\delta > \delta_{\rm c,2}
\end{equation}
hold true. For numerical simulations, we set $\theta_1=5$, $\theta_{12}=8$, and $\delta=0.4$, yielding $\lambda_1=-\frac{1}{3}$ and $\lambda_{2, 3} = -0.014 \pm 0.437i$. Two trajectories, shown in Fig.~\ref{fig:traj}(d), spiral into the presently discussed face equilibrium.

\subsection{Movement of stable equilibria as $\delta$ varies}

\begin{figure}[t!]
\centering 
\includegraphics[width=90mm,trim={0mm 0mm 0mm 0mm}, clip]{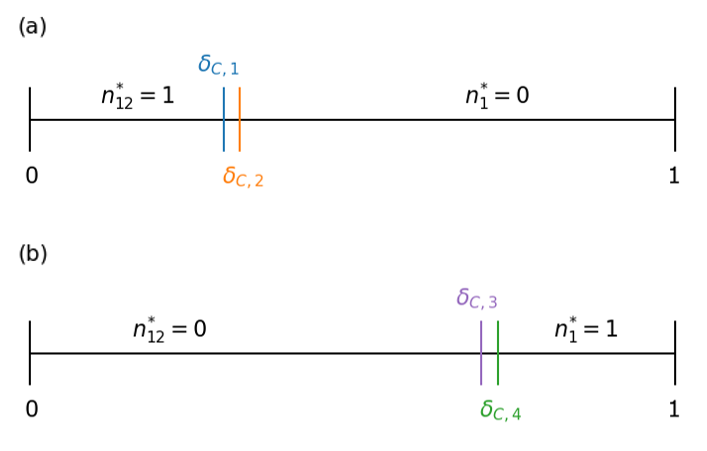}
\caption{Impact of the inter-community interaction rate $\delta$ on stability. Stable edge and face equilibria when $\theta_1 \neq \theta_{12}$ are shown as a function of $\delta$. In both (a) and (b), we use the payoff values given by Eq.~\eqref{nummatri}. (a) $\theta_1 < \theta_{12}$. 
The face equilibrium with $n_{12}^\ast=1$ is stable for $\delta<\delta_{\rm c,1}$.
The edge equilibrium $(x_1^\ast, n_1^\ast, n_{12}^\ast) = \left(\frac{P_0-S_0-\delta(P_0-P_1-S_0+S_1)}{R_0-T_0-S_0+P_0-\delta\gamma},0,1\right)$ is stable for $\delta_{\rm c,1}<\delta<\delta_{\rm c,2}$.
The face equilibrium with $n_1^\ast=0$ is stable for $\delta>\delta_{\rm c,2}$.
(b) $\theta_1 > \theta_{12}$.
The face equilibrium with $n_{12}^\ast=0$ is stable for $\delta<\delta_{\rm c,3}$.
The edge equilibrium $(x_1^\ast, n_1^\ast, n_{12}^\ast) = \left(\frac{P_1-S_1+\delta(P_0-P_1-S_0+S_1)}{R_1-T_1-S_1+P_1+\delta\gamma},1,0\right)$ is stable for $\delta_{\rm c,3}<\delta<\delta_{\rm c,4}$.
The face equilibrium with $n_1^\ast=1$ is stable for $\delta>\delta_{\rm c,4}$. 
In (a), we set $\theta_1 = 5$ and $\theta_{12} = 8$, yielding $\delta_{\rm c,1}=7/22$ and $\delta_{\rm c,2}=10/21$.
In (b), we set $\theta_1 = 8$ and $\theta_{12}=5$, yielding $\delta_{\rm c,3}=21/31$ and $\delta_{\rm c,4} = 15/22$.}
\label{fig:line}
\end{figure}

\begin{figure}[!t]
\centering 
\includegraphics[width=120mm]{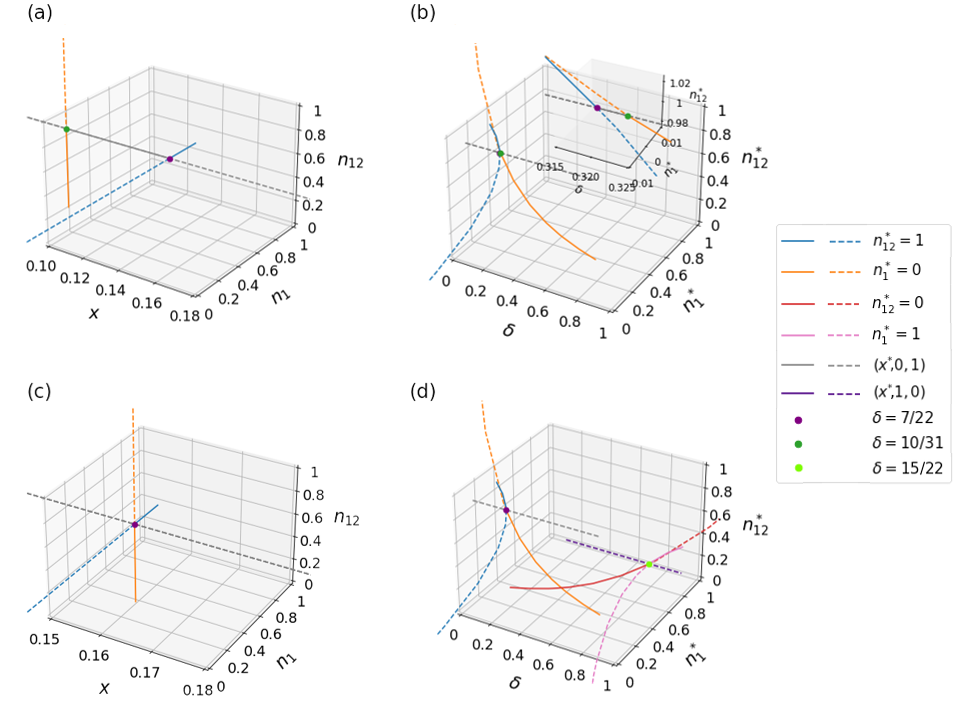} 
\caption{Visualization of the transcritical bifurcations as $\delta$ varies. We use the payoff matrices given by Eq.~\eqref{nummatri}. 
The solid and dashed lines indicate stable and unstable equilibria, respectively, both disregarding the $0$ eigenvalues along the direction of $L$ in the case of $\theta_1 = \theta_{12}$. (a) Movement of three equilibria in the full state space as $\delta$ varies when $\theta_1 = 5$ and $\theta_{12} = 8$. A transcritical bifurcation occurs involving the face equilibrium on $n_{12}=1$ and the edge equilibrium $\left(x^\ast,0,1\right)$, where $x^\ast = 1/6$, at $\delta=7/22$. The second transcritical bifurcation occurs involving the face equilibrium on $n_1=0$ and the edge equilibirium $\left(x^\ast,0,1\right)$, where $x^\ast = 1/9$, at $\delta=10/31$.
(b) Positions of all the same three edge and face equilibria as a function of $\delta$. The $\theta_1$ and $\theta_{12}$ values are the same as those used in (a). In (b), the three curves do not meet at a single point, as shown in the inset, which is a magnification of the main panel.
(c) Same as (a) but when $\theta_1=\theta_{12} = 5$.  A transcritical bifurcation occurs involving the face equilibrium on $n_{12}=1$ and that on $n_1=0$ at $\left(\frac{1}{6},0,1\right)$ when $\delta=7/22$. Edge equilibrium $\left(x^\ast, 0, 1\right)$ also collides with the two face equilibria at this value of $\delta$. 
(d) Same as (b) but when $\theta_1=\theta_{12} = 5$. There is another triplet of equilibria in addition to the triplet of equilibria shown in (c). For this second set of triplet of equilibria, a transcritical bifurcation occurs involving the face equilibrium on $n_{12}=0$ and that on $n_1=1$, and edge equilibrium $\left(x^\ast,1,0\right)$ collides with the bifurcation point, at $\delta=15/22$. 
Note that $x^\ast$ is not constant along the trajectories in (b), whereas it is in (d).
}
\label{fig:bifdia}
\end{figure}

The results in sections~\ref{sec:C1}--\ref{sec:F1} indicate that, for given $\theta_1$ and $\theta_{12}$ ($\neq \theta_1$) values, there are
three equilibria, two of which are face equilibria and one is an edge equilibrium. Just one of these three equilibria is stable for a given value of $\delta$.

Specifically, when $\theta_1 < \theta_{12}$, a face equilibrium is stable when $0 < \delta < \delta_{\rm c, 1}$, an edge equilibrium is stable when $\delta_{\rm c, 1} < \delta < \delta_{\rm c, 2}$, and another face equilibrium is stable when $\delta_{\rm c, 2} < \delta < 1$; see Fig.~\ref{fig:line}(a). As $\delta$ varies, the position of the stable equilibrium continuously moves, including through $\delta = \delta_{\rm c, 1}$ and $\delta = \delta_{\rm c, 2}$.
The dynamical system undergoes a transcritical bifurcation at $\delta=\delta_{\rm c,1}$, with which the face equilibrium and the edge equilibrium exchange the stability. Another similar transcritical bifurcation occurs at $\delta=\delta_{\rm c,2}$. See Figs.~\ref{fig:bifdia}(a) and \ref{fig:bifdia}(b) for visualization.
When $\theta_1 > \theta_{12}$, a different set of three equilibria, which reside on the opposite side of the unit-cube state space, are stable for a respective range of $\delta$, as shown in
Fig.~\ref{fig:line}(b). Similarly to the case of $\theta_1 < \theta_{12}$, these equilibria undergo transcritical bifurcations at $\delta = \delta_{\rm c,3}$ and $\delta_{\rm c,4}$. 

\begin{figure}[t!]
\centering 
\includegraphics[width=120mm]{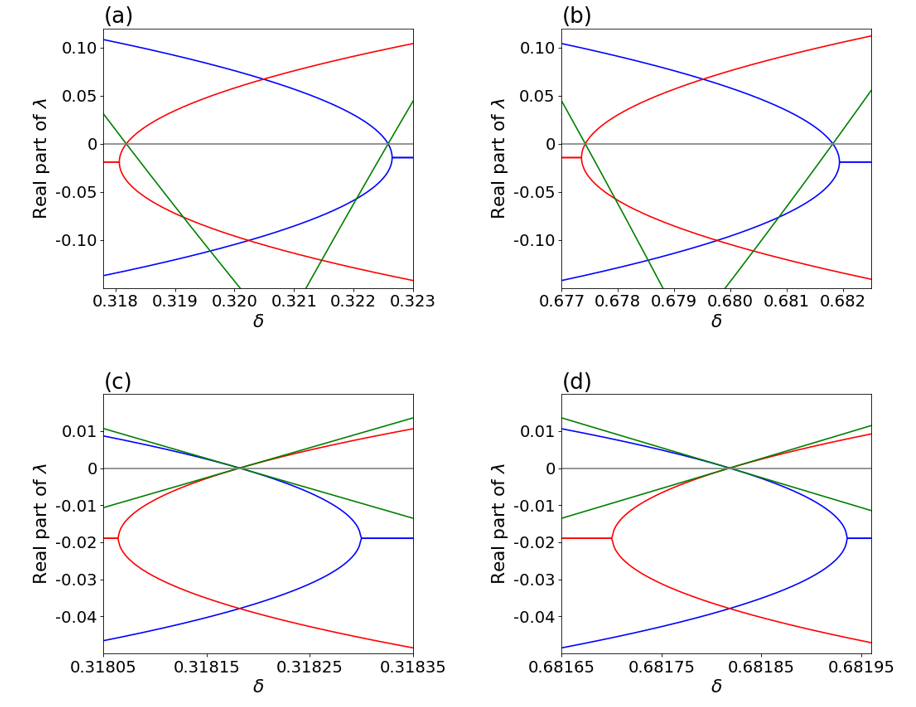} 
\caption{Real part of the eigenvalues of the Jacobian near transcritical bifurcations as a function of $\delta$. We use the payoff matrices given by Eq. \eqref{nummatri}. It should be noted that the third eigenvalue in (a) and (b) is always negative and thus is not shown, and that the third eigenvalue in (c) and (d) is always $0$. (a) $\theta_1=5$ and $\theta_{12}=8$. Each color represents a face or edge equilibrium. Two eigenvalues become $0$ at $\delta = \delta_{\rm c, 1} = 7/22 \approx 0.31818$, and another two eigenvalues become $0$ at $\delta = \delta_{\rm c, 2} = 10/31 \approx 0.32258$. Each of these $\delta$ values marks a transcritical bifurcation. At $\delta \approx 0.31822$ and $0.32321$, the eigenvalues of the stable face equilibrium turns from real to imaginary and vice versa.  
(b) $\theta_1=8$ and $\theta_{12}=5$. Two eigenvalues become $0$ at $\delta = \delta_{\rm c, 3} = 21/31 \approx 0.67742$, and another two eigenvalues become $0$ at $\delta = \delta_{\rm c, 4} = 15/22 \approx 0.68182$. Each of these $\delta$ values marks a transcritical bifurcation. At $\delta \approx 0.67732$ and $0.68194$, the eigenvalues of the stable face equilibrium turns from real to imaginary and vice versa.  
(c) $\theta_1 = \theta_{12} = 5$ and near the first transcritical bifurcation at $\delta = \delta_{\rm c, 1} = 7/22 \approx 0.31818$. At $\delta \approx 0.31806$ and $0.31830$, the eigenvalues of the stable face equilibrium turns from real to imaginary and vice versa.
(d) $\theta_1 = \theta_{12} = 5$ and near the second transcritical bifurcation at $\delta = \delta_{\rm c, 3} = 15/22 \approx 0.68182$. At $\delta \approx 0.68170$ and $0.68194$, the eigenvalues of the stable face equilibrium turns from real to imaginary and vice versa. }
\label{fig:eigenvalc}
\end{figure}

We point out that, as the transcritical bifurcation is approached as $\delta$ gradually increases from $0$, the two eigenvalues are both first complex conjugates with negative real parts and then change to  real negative values. Figure \ref{fig:eigenvalc}(a)
shows the dependence of the real part of the two eigenvalues on $\delta$ around $\delta = \delta_{\rm c,1}$.  When the stable face equilibrium approaches an edge of the unit cube, it becomes a sink, enabling the transcritical bifurcation on the edge. The dependence of the Jacobian eigenvalues of the three equilibria near $\delta = \delta_{\rm c, 2}$ is qualitatively the same as that near $\delta = \delta_{\rm c, 1}$ (see Fig.~\ref{fig:eigenvalc}(b)).

\section{Three-dimensional system with $\theta_1=\theta_{12}$} \label{sec:3dim2}

In this section, as in section~\ref{sec:3dim1}, we assume that $\theta_1=\theta_2$ and that the initial condition satisfies $x=y$ and $n_1=n_2$. Then, $x=y$ and $n_1=n_2$ hold true for any $t>0$. We now further assume that $\theta_1=\theta_2=\theta_{12}$. 
%

\subsection{Corner equilibria}\label{sec:C2}

By setting $x^\ast$, $n_1^\ast$, and $n_{12}^\ast$ to $0$ or $1$, we obtain eight corner equilibria. Similar to the case of 
$\theta_1 \neq \theta_{12}$ (see section~\ref{sec:C1}), each corner equilibrium is a saddle. See Appendix \ref{app:C2} for the proof.

\subsection{Interior equilibria}\label{sec:I2}

In this section, we look for equilibria in the interior of the unit cube, i.e., those satisfying $0 < x^\ast, n_1^\ast, n_{12}^\ast < 1$. By setting $\dot{n}_1 = 0$ and $\dot{n}_{12}=0$ in Eqs. \eqref{3n1} and \eqref{3n12}, respectively, with $\theta_1 = \theta_{12}$, and imposing $n_1^\ast, n_{12}^\ast \notin \{ 0, 1 \}$, we obtain 
\begin{equation}
x^\ast=\frac{1}{1+\theta_1}.
\label{eq:x^*-equal-interior}
\end{equation}
By substituting Eq.~\eqref{eq:x^*-equal-interior} in Eq. \eqref{3x} and imposing $\dot{x}=0$, we obtain
\begin{equation} \label{degline}
n_1^\ast(1-\delta)+n_{12}^\ast\delta=\frac{R_0-T_0-P_0\theta_1+S_0\theta_1}{\rho_1}.
\end{equation}
Any point on this line is an equilibrium. We call Eq.~\eqref{degline} the line of equilibria and denote it by $L$; it is the equilibrium manifold.

\begin{figure}[t]
\centering 
\includegraphics[width=80mm,trim={0mm 0mm 0mm 0mm}, clip]{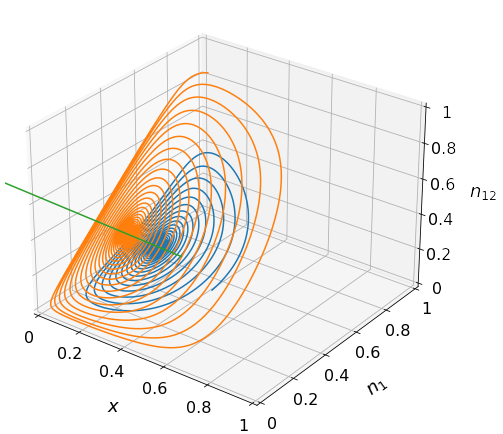} 
\caption{System's behavior near the equilibrium manifold $L$. Shown are trajectories of the three dimensional system given by Eqs.~\eqref{3x}, \eqref{3n1}, and \eqref{3n12} when $\theta_1 = \theta_{12}$ = 5 for two initial conditions. The green line indicates $L$, the line of equilibria given by Eq. \eqref{degline}. We use the payoff matrices given by Eq. \eqref{nummatri}, initial conditions $(x,n_1,n_{12})=(0.5,0.4,0.1)$, shown in blue, and $(0.1,0.9,0.9)$, shown in orange, and set $\delta=0.5$.}
\label{fig:degencase}
\end{figure}

We show in Appendix~\ref{app:I2} that $L$ is neutrally stable along the direction of $L$ and that the other two eigenvalues, $\lambda_2$ and $\lambda_3$, have negative real part 
if Eqs.~\eqref{inequal} and \eqref{stabineq} hold true. In this case, line $L$ attracts trajectories near $L$.

To demonstrate $L$, we numerically simulate trajectories with $\theta_1=5$ and $\delta=0.5$, for which $\lambda_{2, 3} = -0.019 \pm 0.810i$. We show trajectories of the dynamics starting from two initial conditions in Fig.~\ref{fig:degencase}. The figure indicates that the solution spirals into $L$ as expected.

\subsection{Edge equilibria}\label{sec:E2}

Let us examine possible edge equilibria. It should be noted that $\rho_1=\rho_{12}$ when
$\theta_1=\theta_{12}$; we recall that $\rho_1$ and $\rho_{12}$ are defined in Eqs.~\eqref{r1} and \eqref{r12}, respectively. We find that there are just two edge equilibria when $\theta_1 = \theta_{12}$, which are the same as those found for the case $\theta_1 \neq \theta_{12}$ in section~\ref{sec:E1}. These two edge equilibria occur where line $L$ intersects the edge specified by $n_1^\ast=0, n_{12}^\ast=1$ or that specified by $n_1^\ast=1, n_{12}^\ast=0$.

We show in Appendix~\ref{app:E2} that the edge equilibrium $(x^\ast, n_1^\ast, n_{12}^\ast)=\left(\frac{P_0-S_0-\delta(P_0-P_1-S_0+S_1)}{R_0-T_0-S_0+P_0-\delta\gamma},0,1\right)$ is marginally stable with two zero eigenvalues and one negative eigenvalue if and only if Eq. \eqref{stabineq} holds true and
\begin{equation} \label{0,1stabcon}
\delta=\delta_{\rm c, 1} = \delta_{\rm c, 2}.
\end{equation}
 When $\delta \neq \delta_{\rm c,1}$, the Jacobian has two positive eigenvalues and one negative eigenvalue.
Similarly, the edge equilibrium $(x^\ast, n_1^\ast, n_{12}^\ast)=\left(\frac{P_1-S_1+\delta(P_0-P_1-S_0+S_1)}{R_1-T_1-S_1+P_1+\delta\gamma},1,0\right)$ is marginally stable if and only if Eq. \eqref{stabineq} holds true and
\begin{equation} \label{1,0stabcon}
\delta= \delta_{\rm c, 3} = \delta_{\rm c, 4}.
\end{equation}
When $\delta \neq \delta_{\rm c, 3}$, the Jacobian has two positive eigenvalues and one negative eigenvalue.

\subsection{Face equilibria}\label{sec:F2}

\begin{figure}[b!]
\centering 
\includegraphics[width=80mm]{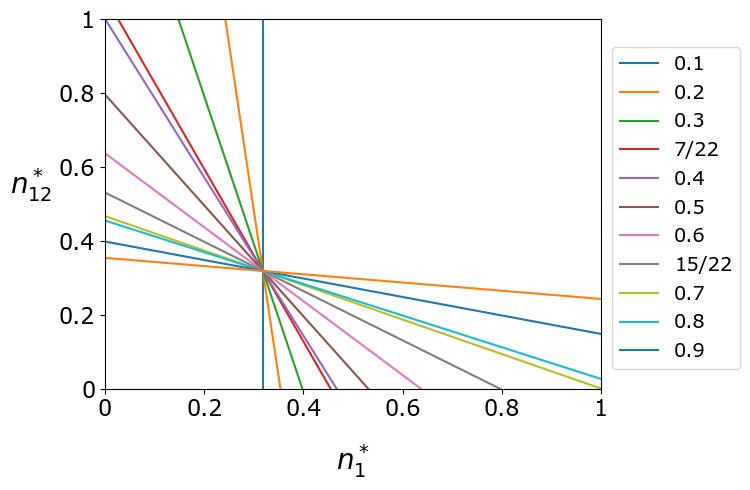} 
\caption{Existence of an invariant point on the line of equilibria, $L$, given by Eq. \eqref{degline} for various values for $\delta$. The legend shows the value of $\delta$ for each line. We use the payoff matrices given by Eq. \eqref{nummatri} and set $\theta_1 = \theta_{12} = 5$.
Because $x^\ast=1/6$, we take the slice of the unit cube with $x^\ast=1/6$ to examine the position of $L$ as a function of $n_1^*$ and $n_{12}^*$. Line $L$ intercepts the point $\left(\frac{1}{6},0,1\right)$ when $\delta = \delta_{\rm c,1} = 7/22$ and the point $\left(\frac{1}{6},1,0\right)$ when $\delta = \delta_{\rm c,3} = 15/22$. All the lines cross at $\left(\frac{1}{6},\frac{7}{22},\frac{7}{22}\right)$, which owes to Eq.~\eqref{eq:focal-point-on-L}.
}
\label{fig:del changes}
\end{figure}

Similarly to the case of the edge equilibria, if we let $x^\ast=0$ or $1$, then we obtain a corner equilibrium.  Therefore, we assume that $0<x^\ast<1$. By setting just one of $n_1^\ast$ or $n_{12}^\ast$ to $0$ or $1$, we obtain the four face equilibria shown in Table~\ref{3dimface1} but with $\theta_1 = \theta_{12}$. Therefore, $x^* = \frac{1}{1+\theta_1}$ for any face equilibria. These face equilibria are stable under the same conditions as those found in section~\ref{sec:F1}, i.e., Eq. \eqref{stabineq}, and the conditions for $\delta$ given by Eqs. \eqref{deltan120}, \eqref{deltan11}, \eqref{deltan121}, and \eqref{deltan10}, i.e., $\delta<\delta_{\rm c,3}$, $\delta>\delta_{\rm c,4} (= \delta_{\rm c, 3})$, $\delta<\delta_{\rm c,1}$, and $\delta>\delta_{\rm c,2} (= \delta_{\rm c, 1})$, respectively.  We also find that these stability requirements for $\delta$ coincide with the requirements for the face equilibria to exist. For example, line $L$ intersects the $n_{12}=0$ face of the unit cube defined by $0\le x, n_1, n_{12} \le 1$ if and only if $\delta$ satisfies Eq.~\eqref{deltan120}, i.e., $\delta<\delta_{\rm c,3}$.

To understand the location of the face equilibria depending on the value of $\delta$, we examine the movement of line $L$ on the ($n_1$, $n_{12}$) plane as we vary $\delta$. The two intersections of $L$ with the boundary of the square defined by $0 \le n_1, n_{12} \le 1$, combined with $x^* = \frac{1}{1+\theta_1}$, give the two face equilibria. When the intersection is at a corner of the square, it is an edge equilibrium. We show $L$ as a function of $\delta$ in Fig.~\ref{fig:del changes} for the payoff matrices given by Eq.~\eqref{nummatri}. Figure~\ref{fig:del changes} indicates that the two edge equilibria are realized at different $\delta$ values, which is consistent with the results shown in section~\ref{sec:E2}.
The figure also indicates that $L$ passes through a particular point regardless of the $\delta$ value. By setting
both the coefficient of $\delta$ and the constant term to $0$ in Eq.~\eqref{degline}, we obtain this point as follows:
\begin{equation}
(x^\ast, n_1^\ast, n_{12}^\ast)=\left(\frac{1}{1+\theta_1}, \frac{R_0-T_0-P_0\theta_1+S_0\theta_1}{\rho_1}, \frac{R_0-T_0-P_0\theta_1+S_0\theta_1}{\rho_1}\right).
\label{eq:focal-point-on-L}
\end{equation}
Figure~\ref{fig:del changes} also indicates that, when $\delta$ is small, $n_1^\ast$ is highly variable between 0 and 1, but the range of $n_{12}^\ast$ is small. When $\delta$ is large, the converse is true. This result is natural because a larger $\delta$ implies that more interaction between players occur between the two communities than in the same community.

As $\delta$ varies, our three-dimensional dynamical system undergoes two bifurcations at $\delta = \delta_{\rm c,1}$ and $\delta = \delta_{\rm c,3}$. When $0<\delta<\delta_{\rm c,1}$, the face equilibrium with $n_{12}^\ast=1$ is stable except along the direction of $L$ (therefore, the Jacobian has two negative eigenvalues and one $0$ eigenvalue), and the edge equilibrium given by $(x^\ast, n_1^\ast, n_{12}^\ast)=\left(\frac{P_0-S_0-\delta(P_0-P_1-S_0+S_1)}{R_0-T_0-S_0+P_0-\delta\gamma},0,1\right)$ and the face equilibrium with $n_1^\ast=0$ are saddles (when disregarding the $0$ eigenvalue along the direction of the line of equilibria; same in the following text). When $\delta=\delta_{\rm c,1}$, the dynamical system undergoes a transcritical bifurcation and the stability of the two face equilibria switches. At $\delta=\delta_{\rm c,1}$, the edge equilibrium has two $0$ eigenvalues and one negative eigenvalue.  These three equilibria collide at $\delta=\delta_{\rm c,1}$, which we depict in Fig.~\ref{fig:bifdia}(c) and (d). When $\delta_{\rm c,1}<\delta<1$, the face equilibrium with $n_{12}^\ast=1$ and the edge equilibrium given by $(x^\ast, n_1^\ast, n_{12}^\ast)=\left(\frac{P_0-S_0-\delta(P_0-P_1-S_0+S_1)}{R_0-T_0-S_0+P_0-\delta\gamma},0,1\right)$ are saddles, and the face equilibrium with $n_1^\ast=0$ is stable. There are three other equilibria located at the other end of $L$ intersecting a face or edge of the state space, i.e., the unit cube.
The structure of the bifurcation occurring at $\delta = \delta_{c, 3}$, involving this second triplet of equilibria, which are composed of two face equilibria (one with $n_1^\ast=1$ and the other with $n_{12}^\ast=0$) and one edge equilibrium given by $(x^\ast, n_1^\ast, n_{12}^\ast)=\left(\frac{P_1-S_1+\delta(P_0-P_1-S_0+S_1)}{R_1-T_1-S_1+P_1+\delta\gamma},1,0\right)$, is qualitatively the same.

Similarly to when $\theta_1 \neq \theta_{12}$, as $\delta$ gradually increases from $0$ to approach the first transcritical bifurcation, the two eigenvalues except the $0$ eigenvalue are first complex conjugates with negative real parts and then change to real negative values. Figure \ref{fig:eigenvalc}(c) shows the dependence of the real part of the two eigenvalues on $\delta$ around $\delta = \delta_{\rm c,1}$. Therefore, when $L$ intersects the unit cube at a point not close to an edge, trajectories on the face spiral into the stable face equilibria, which is consistent with the numerical results shown in Fig.~\ref{fig:degencase}. When the stable face equilibrium approaches an edge of the unit cube, it becomes a sink, enabling the transcritical bifurcation on the edge. The dependence of the Jacobian eigenvalues of the three equilibria near $\delta = \delta_{\rm c, 3}$ is qualitatively the same as that near $\delta = \delta_{\rm c, 1}$ (see Fig.~\ref{fig:eigenvalc}(d)).

\section{Five-dimensional system}\label{sec:5dim}

\begin{figure}[b!]
\centering 
\includegraphics[width=150mm]{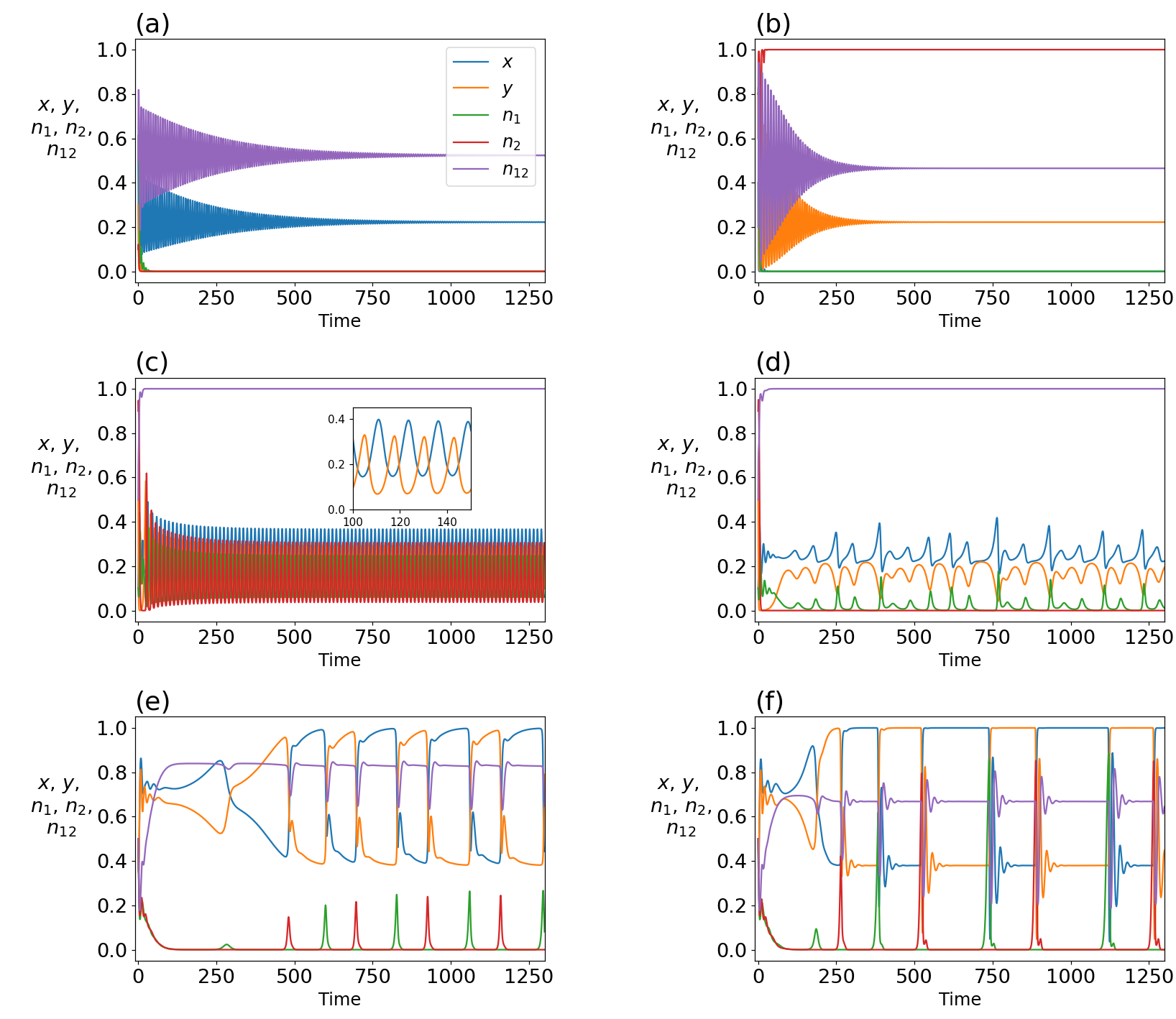} 
\caption{Rich dynamical behavior of the full model. Shown are time courses of trajectories of the five-dimensional system for different parameter choices and initial conditions. 
(a) $\theta_1=3$, $\theta_2=5$, $\theta_{12}=8$, and $\delta=0.95$ with initial condition $(x, y, n_1, n_2, n_{12})=(0.5,0.3,0.5,0.1,0.5)$.
(b) Same parameter values as (a) but with initial condition $(x, y, n_1, n_2, n_{12})=(0.4,0.8,0.8,0.6,0.2)$.
(c) $\theta_1=3$, $\theta_2=5$, $\theta_{12}=8$, and $\delta=0.31$ with initial condition $(x, y, n_1, n_2, n_{12})=(0.1,0.5,0.1,0.9,0.5)$.
(d) $\theta_1=3$, $\theta_2=5$, $\theta_{12}=8$, and $\delta=0.4$ with initial condition $(x, y, n_1, n_2, n_{12})=(0.1,0.5,0.1,0.9,0.5)$.
(e) $\theta_1=0.3$, $\theta_2=0.4$, $\theta_{12}=0.45$, and $\delta=0.29$ with initial condition $(x, y, n_1, n_2, n_{12})=(0.5,0.5,0.5,0.5,0.5)$.
(f) $\theta_1=0.3$, $\theta_2=0.4$, $\theta_{12}=0.45$, and $\delta=0.35$ with initial condition $(x, y, n_1, n_2, n_{12})=(0.5,0.5,0.5,0.5,0.5)$.}
\label{fig:fivedimen}
\end{figure}

In this section, we analyze the five-dimensional dynamical system given by Eqs.~\eqref{n1}--%
%
%
\eqref{5y} without assuming symmetry between the two communities. We exhaustively examine its equilibria as follows.  First, we search for all possible combinations of $x, y, n_1, n_2$, and $n_{12}$ by classifying the value of each variable to be either $0$, $1$, or between $0$ and $1$.  Because three options are available for each variable, there are $3^5=243$ possible combinations.
Second, we find that the $2^5 = 32$ corners of the state space given by $x, y, n_1, n_2, n_{12} \in \{ 0, 1 \}$ are equilibria, more specifically, saddles. Third, out of the remaining $211$ combinations, we have found that $60$ combinations are equilibria; the other $151$ combinations are not. We show these equilibria in Appendix~\ref{app:ST}.
By analyzing the Jacobian of the $60$ equilibria with the assistance of Mathematica, we find that $21$ of them are stable under some conditions (see Appendix~\ref{app:ST}).

In contrast to the reduced three-dimensional dynamical system, there is multistability in the present five-dimensional dynamical system. There are $11$ multistable pairs of equilibria, and these equilibria tend to be multistable when $\delta$ is large. Six of these $11$ pairs are multistable for any $\delta>0.75$. The other five pairs require $\delta$ to be larger, approximately $\delta>0.9$. We demonstrate a multistable pair of equilibria in Figs.~\ref{fig:fivedimen}(a) and (b), which show two trajectories for $\theta_1=3$, $\theta_2=5$, $\theta_{12}=8$, and $\delta=0.95$. The initial condition is $(x, y, n_1, n_2, n_{12})=(0.5,0.3,0.5,0.1,0.5)$ in Fig.~\ref{fig:fivedimen}(a) and $(x, y, n_1, n_2, n_{12})=(0.4,0.8,0.8,0.6,0.2)$ in Fig.~\ref{fig:fivedimen}(b). The trajectory converges towards $(x^\ast, y^\ast, n_1^\ast, n_2^\ast, n_{12}^\ast) \approx (0.222, 0, 0, 0, 0.523)$ in Fig.~\ref{fig:fivedimen}(a) and $\approx (0, 0.222, 0, 1, 0.450)$ in Fig.~\ref{fig:fivedimen}(b).

Figure \ref{fig:fivedimen}(c) shows an oscillatory trajectory for $\theta_1=3$, $\theta_2=5$, $\theta_{12}=8$, $\delta=0.31$, and initial condition $(x, y, n_1, n_2, n_{12})=(0.1,0.5,0.1,0.9,0.5)$. The inset of the figure, showing the time courses of $x$ and $y$, indicates anti-synchronization behavior during the oscillatory dynamics. 
We point out the environmental state between the two communities is bountiful (i.e., $n_{12} \approx 1$) and almost constant despite the anti-synchronous dynamics between $x$ and $y$.
When one increases $\delta$ to $\delta = 0.4$, with all the other parameter values being the same as those used in Fig.~\ref{fig:fivedimen}(c), the oscillations become apparently aperiodic while keeping anti-synchronous behavior between $x$ and $y$
(see Fig.~\ref{fig:fivedimen}(d)). We observe $n_{12} \approx 1$ and $n_2 \approx 0$ during this apparently aperiodic dynamics. It should be noted that $n_1$ is similarly aperiodic.

We show in Fig.~\ref{fig:fivedimen}(e) the trajectory for $\theta_1=0.3$, $\theta_2=0.4$, $\theta_{12}=0.45$, $\delta=0.29$, and initial condition $(x, y, n_1, n_2, n_{12})=(0.5,0.5,0.5,0.5,0.5)$. Similar to Fig.~\ref{fig:fivedimen}(c), the trajectory shown in Fig.~\ref{fig:fivedimen}(e) shows apparent convergence to a limit cycle and approximate anti-synchronization between $x$ and $y$, but accompanying sudden jumps in various variables in each cycle. When $\delta$ is increased to $0.35$, the amplitude of oscillation becomes larger, in particular in terms of $n_1$ and $n_2$ (see Fig.~\ref{fig:fivedimen}(f)).

\section{Discussion}\label{sec:dis}

We extended a previously proposed model of eco-evolutionary dynamics \cite{Weitz16} to the case of networks with two equally sized communities. In the three-dimensional dynamical system given by Eqs. \eqref{3x}, \eqref{3n1}, and \eqref{3n12}, which assumes symmetry between the two communities, a further assumption that $n_1=n_{12}$ lends the model the same as the original well-mixed population model \cite{Weitz16}, and the requirement for the stability of equilibria, i.e., Eq.~\eqref{stabineq}, is the same as that derived in \cite{Weitz16} as well.

Under the generic condition $n_1\neq n_{12}$, our stability requirement for the equilibria again contained that of \cite{Weitz16}, i.e., Eq. \eqref{stabineq}.  However, the stability of the equilibria in our model also requires conditions on the edge weight between two communities, i.e., $\delta$, and on environment recovery rates, i.e., $\theta_1$ ($= \theta_2$) and $\theta_{12}$. When $\theta_1=\theta_{12}$, the line of equilibria, $L$, only requires Eq.~\eqref{stabineq} for stability, but the position of $L$ depends on $\theta_1$ and $\delta$. This result implies that the network has no effect on the stability requirements when $\theta_1=\theta_{12}$. In contrast, when $\theta_1\neq\theta_{12}$, the network and the environment recovery rates affect the stability of the system. As a remark, it was mathematically found \cite{Gong22} that the eco-evolutionary dynamical system proposed in \cite{Weitz16} has no limit cycles. This mathematical result corroborates with the theoretical results in \cite{Weitz16}, in which it was proven that the oscillations converge to a heteroclinic cycle, and our numerical results; because we have analytically shown that there is no internal unstable equilibrium, it is unlikely that our system has a limit cycle. 

There exists another commonly explored family of dynamic payoff matrices dependent on environmental feedback, given by 
\begin{equation}\label{othermatrix}
A(n)=(1-n)\begin{pmatrix}T&P\\R&S\end{pmatrix}+n\begin{pmatrix}R&S\\T&P\end{pmatrix},
\end{equation}
where $T>R$ and $P>S$ \cite{Weitz16, Gong18, Kawano19, Liu20, Stella22, Zhang23}. With Eq.~\eqref{othermatrix}, we retain mutual cooperation as a Nash equilibrium when $n=0$ and mutual defection when $n=1$. In addition, this payoff matrix causes Eq.~\eqref{stabineq} to be satisfied with equality. By using this payoff matrix and holding the assumption that $\theta_1=\theta_{12}$, it is straightforward to analytically obtain a neutrally stable interior line of equilibria, which implies closed periodic orbits in the interior of the state space, corroborating the results in \cite{Weitz16}. 
When $\theta_1\neq\theta_{12}$, our system with Eq.~\eqref{othermatrix} in fact shows a closed periodic orbit on a face of the hypercubic state space.  Therefore, we claim that the closed periodic orbits found in the previous studies with Eq.~\eqref{othermatrix} are at least partially due to the symmetry in the payoff matrix given by Eq.~\eqref{othermatrix}. In the absence of such a symmetry, our results suggest that convergence to stable equilibria is a norm regardless of the population structure. 

When we removed the assumption of symmetry between the two communities by allowing $\theta_1 \neq \theta_2$, we obtained a rich repertoire of stable equilibria, some of which coexist to realize multistability, especially when $\delta$ is large. Multistability was also found in other eco-evolutionary models \cite{Tilman20, Bairagya21}, but these models are ecological extensions of \cite{Weitz16} and are not network-based models as our model is. 
Bistability was also found in a spatial eco-evolutionary model \cite{Hauert19}, but for the trivial equilibria (i.e., bistability between an equilibrium with no cooperators in a replete environment and an equilibrium only with cooperators in a rich environment) and under the snowdrift game. In contrast to these previous studies showing multistability in eco-evolutionary game dynamics, our model is a direct network extension of the original model proposed in \cite{Weitz16} and without additional ecological assumptions. The present results suggest that multistability may be commonly found in the same eco-evolutionary model on various networks. We also found anti-synchronization behavior during oscillatory population dynamics. This type of behavior was found in a prior complete bipartite graph model \cite{Liu20}, but for the division of labor game rather than the typical prisoner's dilemma game. When our stability requirements are not satisfied, our system may converge to a heteroclinic cycle. Further exploring different types of oscillatory behavior in networked eco-evolutionary game dynamics may be interesting.

We emphasize that our model substantially varies from the previously proposed model composed of two interacting subpopulations, or precisely, complete bipartite graphs \cite{Gong18, Kawano19, Fu22}. Their model does not allow interaction between players in the same subpopulation, whereas our model does. Furthermore, these previous studies adopted the dynamic payoff matrix given by Eq.~\eqref{othermatrix}, which led to closed periodic orbits, as we discussed above. In \cite{Gong18, Kawano19}, such cyclic orbits do not accompany anti-synchronous oscillation of the fraction of cooperation in the two subpopulations. Instead, the cyclic behavior originates from interplay of the fraction of one of the two subpopulations and the environmental variable. On the other hand, the orbits obtained in \cite{Fu22} show largely in-phase synchronous oscillation between the two subpopulations. The model in \cite{Gong18, Kawano19} was extended in \cite{Liu20} to include a different form of $A(n)$ and different influences of strategies in two subpopulations on the environment. The inclusion of these parameters produces periodic orbits as did the models proposed in \cite{Gong18, Kawano19}. In contrast, our model showed anti-phase oscillations in terms of the fraction of cooperators in the two communities (i.e., $x$ and $y$) and multistability. Therefore, even within the family of two-subpopulation networks, which is one of the simplest network model, qualitatively different dynamical behavior may arise depending on the assumption on the environmental dynamics.

Prior extensions of the eco-evolutionary game models to larger networks include those to spatial lattices and regular graphs. The spatial extensions have been to the case of square lattices \cite{Jin18, Szolnoki17, Wu18, Hauert19, Lin19, Wu19, Yang21, Ding23, He23, Lu21, Zhu23}. A lattice model of eco-evolutionary game dynamics assuming local environmental variables, meaning that each node (i.e., player) has its own dynamical environmental state, resulted in spatiotemporal patterns, including clustering, flickering, and wave-like patterns \cite{Lin19}.  
Enhanced cooperation due to the environmental feedback was also found in eco-evolutionary models on square lattices \cite{Jin18, Szolnoki17, Wu18, Wu19, Ding23, He23, Lu21,  Zhu23}.
%
%
Another type of network that has been studied with eco-evolutionary feedback is regular graphs, in which all nodes have degree $k$. Through the use of pair approximation, the extension of the original model \cite{Weitz16} to regular graphs
(therefore using the payoff matrix given by Eq.~\eqref{othermatrix}) has clarified that an increased $k$ induces the internal stable equilibrium to become neutrally stable, producing periodic orbits \cite{Stella22, Zhang23}. These models are substantially different from ours not only in the network structure but also in that their model assumes that the environment is global to all nodes. Assigning an environmental state $n_{ij}$ to each edge $(i, j)$, as has been done for square lattices in previous studies \cite{He23, Zhu23} and for a two-community network in the present study, in the case of regular graphs and general networks may be an interesting generalization.

In addition to the extension of the network structure, edge-dependent environmental state variable, and weighted networks, which we discussed above, there are further possible extensions of the present model as future work. First, in well-mixed populations, incorporation of intrinsic environmental dynamics, such as resource growth and decay, results in multistability and limit cycles \cite{Tilman20}, which one can explore for networks. 
Second, the incorporation of dynamic recovery and degradation rates for the environmental state, which are boosted by cooperators’ and defectors' payoffs \cite{Cao21}, leads to the same stability requirement as that in \cite{Weitz16}, i.e., Eq.~\eqref{stabineq}. One can extend the present model to the case of dynamic rates of environment recovery and degradation by letting, e.g., $\theta_1$ depend on $x$ and $n_1$.
Third, the use of finite carrying capacity in an environment, which excludes any periodic orbits and enables bistability in the original model \cite{Bairagya21}, should be possible. Fourth, the incorporation of aspiration dynamics, with which players update their strategies based on whether or not they are satisfied with their current payoff \cite{Arefin21} is another possible direction of research.  Lastly, although we studied the prisoner's dilemma, as other eco-evolutionary game dynamics models, our model can be studied for other games such as the prisoner's dilemma with voluntary participation \cite{Liu20, Li21}, coordination game \cite{Weitz16, Lin19, Bairagya21, Fu22}, anti-coordination game \cite{Weitz16, Lin19, Bairagya21}, and division-of-labor game \cite{Liu20}.

In conclusion, we have studied an eco-evolutionary game dynamics model with two distinct network communities. We find that the interaction rates both within and between these communities significantly impact on the resulting dynamical behavior and the determination of possible equilibrium classes (i.e., interior, face, edge, and corner) of the system. In addition to numerical investigation of the full model, we have performed comprehensive stability analysis of the simplified system under symmetry conditions. Our work highlights the importance of community structures in impacting eco-evolutionary dynamics across different ecological niches.
\section*{Acknowledgments} 
N.M. acknowledges support from the Japan Science and Technology Agency (JST) Moonshot R\&D (under Grant No.\,JPMJMS2021), the National Science Foundation (under Grant Nos.\,2052720 and 2204936), and JSPS KAKENHI (under grant Nos.\,JP 21H04595 and 23H03414).
\bibliographystyle{unsrt}
\bibliography{Betz-Research-Refs}

\begin{appendices}
\section{Corner equilibria of the three-dimensional system with $\theta_1\neq \theta_{12}$}\label{app:C1}

By evaluating Eq. \eqref{3Jac} at each corner equilibrium, we obtain
\begin{align}
J(0,0,0)&=\begin{pmatrix}
S_0-P_0 & 0 & 0\\
0 & -1 & 0\\
0&0&-1
\end{pmatrix},\\
J(0,0,1)&=\begin{pmatrix}
-P_0(1-\delta)-P_1\delta+S_0(1-\delta)+S_1\delta & 0 & 0\\
0 & -1 & 0\\
0&0&1
\end{pmatrix},\\
J(0,1,0)&=\begin{pmatrix}
-P_0\delta-P_1(1-\delta)+S_0\delta+S_1(1-\delta) & 0 & 0\\
0 & 1 & 0\\
0&0&-1
\end{pmatrix},\\
J(0,1,1)&=\begin{pmatrix}
S_1-P_1 & 0 & 0\\
0 & 1 & 0\\
0&0&1
\end{pmatrix},\\
J(1,0,0)&=\begin{pmatrix}
T_0-R_0 & 0 & 0\\
0 & \theta_1 & 0\\
0&0&\theta_{12}
\end{pmatrix},\\
J(1,0,1)&=\begin{pmatrix}
-R_0(1-\delta)-R_1\delta+T_0(1-\delta)+T_1\delta & 0 & 0\\
0 & \theta_1 & 0\\
0&0&-\theta_{12}
\end{pmatrix},\\
J(1,1,0)&=\begin{pmatrix}
-R_0\delta-R_1(1-\delta)+T_0\delta+T_1(1-\delta) & 0 & 0\\
0 & -\theta_1 & 0\\
0&0&\theta_{12}
\end{pmatrix},\\
J(1,1,1)&=\begin{pmatrix}
T_1-R_1 & 0 & 0\\
0 & -\theta_1 & 0\\
0&0&-\theta_{12}
\end{pmatrix}.
\end{align}
By Eq. \eqref{inequal}, we obtain that $S_0-P_0>0$, $T_0-R_0<0$, $S_1-P_1<0$, and $T_1-R_1>0$.  Therefore, each of these Jacobians has at least one positive eigenvalue and one negative eigenvalue, and each corner equilibrium is a saddle.

\section{Edge equilibria of the three-dimensional system with $\theta_1\neq \theta_{12}$} \label{app:E1}

\subsection{Equilibrium $(x^\ast, n_1^\ast, n_{12}^\ast)=\left(\frac{P_0-S_0-\delta(P_0-P_1-S_0+S_1)}{R_0-T_0-S_0+P_0-\delta\gamma},0,1\right)$}\label{app:E11}

For the equilibrium $(x^\ast, n_1^\ast, n_{12}^\ast)=\left(\frac{P_0-S_0-\delta(P_0-P_1-S_0+S_1)}{R_0-T_0-S_0+P_0-\delta\gamma},0,1\right)$, the Jacobian, Eq.~\eqref{3Jac}, is reduced to
\begin{equation}
J = \begin{pmatrix}
J^{(2)}_{11} &J^{(2)}_{12}  & J^{(2)}_{13}\\
0&J^{(2)}_{22}&0\\
0&0&J^{(2)}_{33}
\end{pmatrix},
\end{equation}
where
\begin{align}
J^{(2)}_{11}&=\frac{[(P_0-S_0)(1-\delta)+\delta(P_1-S_1)][(R_0-T_0)(1-\delta)+\delta(R_1-T_1)]}{R_0-T_0-S_0+P_0-\delta\gamma}\label{76},\\
J^{(2)}_{12}&=\frac{(1-\delta)\omega[(P_0-S_0)(1-\delta)+\delta(P_1-S_1)][(R_0-T_0)(1-\delta)+\delta(R_1-T_1)]}{(R_0-T_0-S_0+P_0-\delta\gamma)^3}\label{77},\\
J^{(2)}_{13}&=\frac{\delta\omega[(P_0-S_0)(1-\delta)+\delta(P_1-S_1)][(R_0-T_0)(1-\delta)+\delta(R_1-T_1)]}{(R_0-T_0-S_0+P_0-\delta\gamma)^3}\label{78},\\
J^{(2)}_{22}&=-\frac{R_0-T_0-P_0\theta_1+S_0\theta_1-\delta\rho_1}{R_0-T_0-S_0+P_0-\delta\gamma}\label{79},\\
J^{(2)}_{33}&=\frac{R_0-T_0-P_0\theta_{12}+S_0\theta_{12}-\delta\rho_{12}}{R_0-T_0-S_0+P_0-\delta\gamma},\label{80}
\end{align}
and
\begin{equation}
\omega= (P_0-S_0)(R_1-T_1)-(P_1-S_1)(R_0-T_0)
\label{omeg}.
\end{equation}

The eigenvalues of $J$ are given by
$\lambda = J^{(2)}_{11}, J^{(2)}_{22}, J^{(2)}_{33}$.
Using the fact that each eigenvalue has a common denominator, which we refer to as
\begin{equation}
\mu_1 \equiv R_0-T_0-S_0+P_0-\delta\gamma,
\end{equation}
we determine the stability of the edge equilibrium by examining the following four possible cases.

\subsubsection*{Case 1: $\gamma>0$ and $\mu_1>0$}

First, $\gamma$ is positive if and only if 
\begin{equation}\label{gampos}
R_0-T_0-R_1+T_1>S_0-P_0-S_1+P_1.
\end{equation}
Under this condition, $\mu_1$ is positive if and only if
\begin{equation}\label{m1pos}
\delta<\frac{R_0-T_0-S_0+P_0}{\gamma}.
\end{equation}
For positive $\delta$ values satisfying Eq.~\eqref{m1pos} to exist, it must hold true that $R_0-T_0>S_0-P_0$. 

Eigenvalues $J^{(2)}_{22}$ and $J^{(2)}_{33}$ are negative if and only if 
\begin{align}
\delta&<\frac{R_0-T_0-P_0\theta_1+S_0\theta_1}{\rho_1}\label{150}
\intertext{and}
\delta&> \frac{R_0-T_0-P_0\theta_{12}+S_0\theta_{12}}{\rho_{12}},
\label{151}
\end{align}
respectively. Lastly, eigenvalue $J^{(2)}_{11}$ is negative if either
\begin{align}
\delta< \min \left\{ \frac{P_0-S_0}{P_0-S_0-P_1+S_1}, \frac{R_0-T_0}{R_0-T_0-R_1+T_1} \right\}\label{152}
\end{align}
or
\begin{align}
\delta > \max \left\{ \frac{P_0-S_0}{P_0-S_0-P_1+S_1}, \frac{R_0-T_0}{R_0-T_0-R_1+T_1} \right\}\label{153}
\end{align}
holds true. However, we find that there is no $\delta$ value that simultaneously satisfies Eqs.~\eqref{m1pos}, \eqref{150}, \eqref{151}, and \eqref{152}, or one that simultaneously satisfies Eqs.~\eqref{m1pos}, \eqref{150}, \eqref{151}, and \eqref{153}.

\subsubsection*{Case 2: $\gamma>0$ and $\mu_1 < 0$}

If $\gamma > 0$, then $\mu_1 < 0$ if and only if 
\begin{equation}\label{m1neg}
\delta>\frac{R_0-T_0-S_0+P_0}{\gamma}.
\end{equation}
For positive $\delta$ values satisfying Eq.~\eqref{m1neg} to exist and be less than $1$, it must hold true that $R_1-T_1<S_1-P_1$. Eigenvalues $J^{(2)}_{22}$ and $J^{(2)}_{33}$ are negative if and only if
\begin{align}
\delta>&\frac{R_0-T_0-P_0\theta_1+S_0\theta_1}{\rho_1} = \delta_{\rm c,1}\label{deledge1}
\intertext{and}
\delta<&\frac{R_0-T_0-P_0\theta_{12}+S_0\theta_{12}}{\rho_{12}} = \delta_{\rm c,2},
\label{deledge2}
\end{align}
respectively.
For a $\delta$ value satisfying Eqs.~\eqref{deledge1} and \eqref{deledge2} to exist, it must hold true that $\theta_{12}>\theta_1$.  

Lastly, because we have assumed that $\mu_1 < 0$, the numerator of Eq.~\eqref{76} has to be positive for eigenvalue $J^{(2)}_{11}$ to be negative. Then, either both $(P_0-S_0)(1-\delta)+\delta(P_1-S_1)$ and $(R_0-T_0)(1-\delta)+\delta(R_1-T_1)$ are positive or both are negative.  

If both $(P_0-S_0)(1-\delta)+\delta(P_1-S_1)$ and $(R_0-T_0)(1-\delta)+\delta(R_1-T_1)$ are positive, we obtain
\begin{align}
\delta&>\frac{P_0-S_0}{P_0-S_0-P_1+S_1}\label{160}
\intertext{and}
\delta&<\frac{R_0-T_0}{R_0-T_0-R_1+T_1}.\label{161}
\end{align}
We find that there is no $\delta$ value that simultaneously satisfies Eqs.~\eqref{m1neg}, \eqref{deledge1}, \eqref{deledge2}, \eqref{160}, and \eqref{161}.

If both $(P_0-S_0)(1-\delta)+\delta(P_1-S_1)$ and $(R_0-T_0)(1-\delta)+\delta(R_1-T_1)$ are negative, we obtain
\begin{align}
\delta&<\frac{P_0-S_0}{P_0-S_0-P_1+S_1}\label{deledge11}
\intertext{and}
\delta&>\frac{R_0-T_0}{R_0-T_0-R_1+T_1}.\label{deledge22}
\end{align}
A $\delta$ value satisfying Eqs.~\eqref{deledge11} and \eqref{deledge22} exists if and only if
\begin{equation}
\frac{R_0-T_0}{R_0-T_0-R_1+T_1}<\frac{P_0-S_0}{P_0-S_0-P_1+S_1},
\end{equation}
which is equivalent to Eq.~\eqref{stabineq}.

Such a $\delta$ value satisfying $\delta \in (0, 1)$ exists if and only if 
the right-hand side (RHS) of Eq.~\eqref{deledge11} is positive and that of Eq.~\eqref{deledge22} is less than $1$.
In fact, because $P_0-S_0<0$ and $S_1-P_1<0$, the numerator and denominator on the RHS of Eq.~\eqref{deledge11} are negative, which implies that the RHS is positive. The RHS of Eq.~\eqref{deledge22} is less than $1$ because
we obtain $R_0-T_0 > 0$ and $R_1 - T_1 < 0$ from Eq.~\eqref{inequal}. Lastly, the intersection of the conditions derived for $\delta$, given by Eqs.~\eqref{m1neg}, \eqref{deledge1}, \eqref{deledge2}, \eqref{deledge11}, and \eqref{deledge22}, yields Eq.~\eqref{delline1}, i.e., $\delta_{\rm c,1} <\delta < \delta_{\rm c,2}$.

\subsubsection*{Case 3: $\gamma< 0$ and $\mu_1 > 0$}

In this section, we assume that $\gamma < 0$, i.e
\begin{equation}\label{gamneg}
R_0-T_0-R_1+T_1<S_0-P_0-S_1+P_1.
\end{equation}
Then, $\mu_1 > 0$ if and only if Eq.~\eqref{m1neg} is satisfied. For positive $\delta$ values satisfying Eq.~\eqref{m1neg} to exist and be less than $1$, it must hold true that $R_1-T_1<S_1-P_1$. If $\mu_1 > 0$, then we obtain Eqs.~\eqref{150}, \eqref{151}, \eqref{152}, and \eqref{153}, and again find that there is no $\delta$ value that simultaneously satisfies these inequalities.

\subsubsection*{Case 4: $\gamma<0$ and $\mu_1 < 0$}

In this section, we assume that $\gamma < 0$ and $\mu_1 < 0$, which requires Eq.~\eqref{m1pos} to be satisfied. For positive $\delta$ values satisfying Eq.~\eqref{m1pos} to exist, it must hold true that $R_0-T_0>S_0-P_0$. The derivation of the stability of this case follows the same derivation as Case 2, and we find that the equilibrium is stable if and only if Eqs. \eqref{stabineq}, \eqref{delline1}, and \eqref{t12ltt1} hold true.

\subsection{Equilibrium $(x^\ast, n_1^\ast, n_{12}^\ast)=\left(\frac{P_1-S_1+\delta(P_0-P_1-S_0+S_1)}{R_1-T_1-S_1+P_1+\delta\gamma},1,0\right)$}\label{app:E12}

For the equilibrium $(x^\ast, n_1^\ast, n_{12}^\ast)=\left(\frac{P_1-S_1+\delta(P_0-P_1-S_0+S_1)}{R_1-T_1-S_1+P_1+\delta\gamma},1,0\right)$, the Jacobian is given by
\begin{equation}
J = \begin{pmatrix}
J^{(3)}_{11} &J^{(3)}_{12}  & J^{(3)}_{13}\\
0&J^{(3)}_{22}&0\\
0&0&J^{(3)}_{33}
\end{pmatrix},
\end{equation}
where
\begin{align}
J^{(3)}_{11}&=\frac{[(P_1-S_1)(1-\delta)+\delta(P_0-S_0)][(R_1-T_1)(1-\delta)+\delta(R_0-T_0)]}{R_1-T_1-S_1+P_1+\delta\gamma}\label{81},\\
J^{(3)}_{12}&=\frac{(1-\delta)\omega[(P_1-S_1)(1-\delta)+\delta(P_0-S_0)][(R_1-T_1)(1-\delta)+\delta(R_0-T_0)]}{(R_1-T_1-S_1+P_1+\delta\gamma)^3}\label{82},\\
J^{(3)}_{13}&=\frac{\delta\omega[(P_1-S_1)(1-\delta)+\delta(P_0-S_0)][(R_1-T_1)(1-\delta)+\delta(R_0-T_0)]}{(R_0-T_0-S_0+P_0-\delta\gamma)^3}\label{83},\\
J^{(3)}_{22}&=\frac{R_1-T_1-P_1\theta_1+S_1\theta_1+\delta\rho_1}{R_1-T_1-S_1+P_1+\delta\gamma}\label{84},\\
J^{(3)}_{33}&=-\frac{R_1-T_1-P_1\theta_{12}+S_1\theta_{12}+\delta\rho_{12}}{R_1-T_1-S_1+P_1+\delta\gamma}.\label{85}
\end{align}
The eigenvalues of $J$ are given by
$\lambda = J^{(3)}_{11}, J^{(3)}_{22}, J^{(3)}_{33}$. Similarly to the prior equilibrium, we see that each eigenvalue has a common denominator, which we refer to as
\begin{equation}
\mu_2 \equiv R_1-T_1-S_1+P_1+\delta\gamma.
\end{equation}
We determine the stability of the edge equilibrium by examining the following four possible cases.

\subsubsection*{Case 1: $\gamma>0$ and $\mu_2>0$} 

First, $\gamma$ is positive if and only if Eq.~\eqref{gampos} is satisfied.  Then, $\mu_2$ is positive if and only if
\begin{equation}\label{m2pos}
\delta>\frac{-(R_1-T_1-S_1+P_1)}{\gamma}.
\end{equation}
For positive $\delta$ values satisfying Eq.~\eqref{m2pos} to exist and be less than $1$, it must hold true that $R_0-T_0>S_0-P_0$. 

Eigenvalues $J^{(3)}_{22}$ and $J^{(3)}_{33}$ are negative if and only if 
\begin{align}
\delta&<\frac{-(R_1-T_1-P_1\theta_1+S_1\theta_1)}{\rho_1}\label{162}
\intertext{and}
\delta&> \frac{-(R_1-T_1-P_1\theta_{12}+S_1\theta_{12})}{\rho_{12}},\label{163}
\end{align}
respectively.  Lastly, $J^{(3)}_{11}$ is negative if either
\begin{align}
\delta< \min \left\{ \frac{-(P_1-S_1)}{P_0-S_0-P_1+S_1},\frac{-(R_1-T_1)}{R_0-T_0-R_1+T_1}\right \}\label{164}
\end{align}
or
\begin{align}
\delta > \max \left\{\frac{-(P_1-S_1)}{P_0-S_0-P_1+S_1},\frac{-(R_1-T_1)}{R_0-T_0-R_1+T_1}\right \}\label{165}
\end{align}
holds true.  However, we find that there is no $\delta$ value that simultaneously satisfies Eqs.~\eqref{m2pos}, \eqref{162}, \eqref{163}, and \eqref{164} or Eqs.~\eqref{m2pos}, \eqref{162}, \eqref{163}, and \eqref{165}.

\subsubsection*{Case 2: $\gamma >0$ and $\mu_2<0$} 

If $\gamma$ is positive, then $\mu_2<0$ if and only if 
\begin{equation}\label{m2neg}
\delta<\frac{-(R_1-T_1-S_1+P_1)}{\gamma}.
\end{equation}
For positive $\delta$ values satisfying Eq.~\eqref{m2neg} to exist, it must hold true that $R_1-T_1<S_1-P_1$. Eigenvalues $J^{(3)}_{22}$ and $J^{(3)}_{33}$ are negative if and only if
\begin{equation}\label{deledge3}
\delta> \frac{-R_1+T_1+P_1\theta_1-S_1\theta_1}{\rho_1} = \delta_{\rm c,3}
\end{equation}
and
\begin{equation}\label{deledge4}
\delta< \frac{-R_1+T_1+P_1\theta_{12}-S_1\theta_{12}}{\rho_{12}} = \delta_{\rm c,4},
\end{equation}
respectively. For a $\delta$ value satisfying Eqs.~\eqref{deledge3} and \eqref{deledge4} to exist, it must hold true that $\theta_{12}<\theta_1$.  

Because we have assunmed that $\mu_2<0$, the numerator of Eq.~\eqref{81} has to be positive for eigenvalue $J^{(3)}_{11}$ to be negative.  Then, either both $(P_1-S_1)(1-\delta)+\delta(P_0-S_0)$ and $(R_1-T_1)(1-\delta)+\delta(R_0-T_0)$ are positive or both are negative.  

If both $(P_1-S_1)(1-\delta)+\delta(P_0-S_0)$ and $(R_1-T_1)(1-\delta)+\delta(R_0-T_0)$ are positive, we obtain
\begin{align}
\delta&<\frac{-(P_1-S_1)}{P_0-S_0-P_1+S_1}\label{166}
\intertext{and}
\delta&>\frac{-(R_1-T_1)}{R_0-T_0-R_1+T_1}\label{167}.
\end{align}
We find that there is no $\delta$ value that simultaneously satisfies Eqs.~\eqref{m2neg}, \eqref{deledge3}, \eqref{deledge4}, \eqref{166}, and \eqref{167}.

If both $(P_1-S_1)(1-\delta)+\delta(P_0-S_0)$ and $(R_1-T_1)(1-\delta)+\delta(R_0-T_0)$ are negative, we obtain
\begin{equation}\label{deledge33}
\delta>\frac{-(P_1-S_1)}{P_0-S_0-P_1+S_1}
\end{equation}
and 
\begin{equation}\label{deledge44}
\delta<\frac{-(R_1-T_1)}{R_0-T_0-R_1+T_1}.
\end{equation}
A value of $\delta$ satisfying Eqs. \eqref{deledge33} and \eqref{deledge44} exists if and only if
\begin{equation}
\frac{-(R_1-T_1)}{R_0-T_0-R_1+T_1}<\frac{-(P_1-S_1)}{P_0-S_0-P_1+S_1},
\end{equation}
which is equivalent to Eq. \eqref{stabineq}.  Such a $\delta$ value satisfying $\delta \in (0, 1)$ exists if and only if 
the RHS of Eq.~\eqref{deledge33} is less than $1$ and that of Eq.~\eqref{deledge44} is positive.
In fact, the RHS of Eq.~\eqref{deledge33} is less than $1$ because we obtain $P_0-S_0 < 0$ and $P_1 - S_1 > 0$ from Eq.~\eqref{inequal}. Because $R_0-T_0>0$ and $T_1-R_1>0$, the numerator and denominator on the RHS of Eq.~\eqref{deledge44} are positive, which implies that the RHS is positive. Lastly, the intersection of the conditions derived for $\delta$, given by Eqs.~\eqref{m2neg}, \eqref{deledge3}, \eqref{deledge4}, \eqref{deledge33}, and \eqref{deledge44}, yields Eq.~\eqref{delline2}, i.e., $\delta_{\rm c,3} <\delta < \delta_{\rm c,4}$.

\subsubsection*{Case 3: $\gamma<0$ and $\mu_2>0$} 

In this section, we assume that $\gamma<0$, i.e., Eq.~\eqref{gamneg}. Then, $\mu_2>0$ if and only if Eq.~\eqref{m2neg} is satisfied. For positive $\delta$ values satisfying Eq.~\eqref{m2neg} to exist, it must hold true that $R_1-T_1<S_1-P_1$. If $\mu_2>0$, then we obtain Eqs.~\eqref{162}, \eqref{163}, \eqref{164}, \eqref{165}, and \eqref{m2neg}. However, there is no $\delta$ value that simultaneously satisfies these inequalities.

\subsubsection*{Case 4: $\gamma<0$ and $\mu_2<0$} 

In this section, we assume that $\gamma<0$ and $\mu_2<0$, which requires Eq.~\eqref{m2pos}. For positive $\delta$ values satisfying Eq.~\eqref{m2pos} to exist and be less than $1$, it must hold true that $R_0-T_0>S_0-P_0$. The derivation of the stability of this case follows the same derivation as Case 2, and we find that the equilibrium is stable if and only if Eqs.~\eqref{stabineq}, \eqref{delline2}, and \eqref{t12ltt1} hold true.

\section{Three face equilibria of the three-dimensional system with $\theta_1\neq \theta_{12}$} \label{app:F1}

In this section, we derive the stability conditions for three face equilibria of the three-dimensional system with $\theta_1\neq \theta_{12}$.

\subsection{Equilibrium $(x^\ast, n_1^\ast, n_{12}^\ast) = \left(\frac{1}{1+\theta_{12}},1,\frac{R_1 - T_1 - P_1\theta_{12} + S_1\theta_{12} +\delta\rho_{12}}{\delta \rho_{12}}\right)$}

At $(x^\ast, n_1^\ast, n_{12}^\ast)=\left(\frac{1}{1+\theta_{12}},1,\frac{R_1 - T_1 - P_1\theta_{12} + S_1\theta_{12} +\delta\rho_{12}}{\delta \rho_{12}}\right)$, the Jacobian is reduced to
\begin{equation}\label{3J11}
J=\begin{pmatrix}
J^{(4)}_{11} &J^{(4)}_{12}  & J^{(4)}_{13}\\
0&J^{(4)}_{22}&0\\
J^{(4)}_{31}&0&0
\end{pmatrix},
\end{equation}
where
\begin{align}
J^{(4)}_{11}&=\frac{\left[(P_1 - S_1) (R_0 - T_0) - (P_0 - S_0) (R_1 - T_1)\right]\theta_{12}}{(1 + \theta_{12}) \rho_{12}},\label{79}\\
J^{(4)}_{12}&=\frac{-(1-\delta) \theta_{12} \rho_{12}}{(1 + \theta_{12})^3},\label{81}\\
J^{(4)}_{13}&=\frac{-\delta \theta_{12} \rho_{12}}{(1 + \theta_{12})^3}\label{83},\\
J^{(4)}_{22}&=\frac{\theta_{12}-\theta_1}{\theta_{12}+1}\label{82},\\
J^{(4)}_{31}&=\frac{-(1 + \theta_{12})\left[ R_1 -T_1- (P_1 - S_1)\theta_{12}\right] (R_1 - T_1- P_1\theta_{12} + S_1 \theta_{12}+\delta \rho_{12})}{\delta^2 \rho_{12}^2}\label{80}.
\end{align}
The characteristic equation is given by 
\begin{equation}
\left(J^{(4)}_{22}-\lambda\right)\left(\lambda^2-J^{(4)}_{11}\lambda-J^{(4)}_{13}J^{(4)}_{31}\right)=0\label{3char3}.
\end{equation}
Eigenvalue $\lambda_1 = J^{(4)}_{22} = \frac{\theta_{12}-\theta_1}{\theta_{12}+1}$ is negative if and only if $\theta_{12}<\theta_1$ (i.e., Eq.~\eqref{t12ltt1}).
The real part of the other two eigenvalues is negative if and only if $-J^{(4)}_{11}>0$ and $-J^{(4)}_{13}J^{(4)}_{31}>0$. Equation~\eqref{79} combined with $\rho_{12} > 0$ implies that $-J^{(4)}_{11}>0$ if and only if Eq. \eqref{stabineq} holds true. Because $J^{(4)}_{13} < 0$, condition $-J^{(4)}_{13}J^{(4)}_{31}>0$ is equivalent to $J^{(4)}_{31} > 0$, which holds true if and only if $\delta > \delta_{\rm c,4}$ (i.e., Eq.~\eqref{deltan11}).

In sum, this face equilibrium is stable if and only if Eqs. \eqref{stabineq}, \eqref{t12ltt1}, and \eqref{deltan11} hold true. 

\subsection{Equilibrium $(x^\ast, n_1^\ast, n_{12}^\ast) = \left(\frac{1}{1+\theta_1},\frac{R_0 -T_0- P_0\theta_1 + S_0\theta_1 -\delta\rho_1}{(1- \delta) \rho_1}, 1\right)$}

At $(x^\ast, n_1^\ast, n_{12}^\ast)=\left(\frac{1}{1+\theta_1},\frac{R_0 -T_0- P_0\theta_1 + S_0\theta_1 -\delta\rho_1}{(1- \delta) \rho_1}, 1\right)$, the Jacobian is reduced to
\begin{equation}\label{3J121}
J = \begin{pmatrix}
J^{(5)}_{11} &J^{(5)}_{12}  & J^{(5)}_{13}\\
J^{(5)}_{21}&0&0\\
0&0&J^{(5)}_{33}
\end{pmatrix},
\end{equation}
where 
\begin{align}
J^{(5)}_{11}&=J^{(1)}_{11},\label{62}\\
J^{(5)}_{12}&=J^{(1)}_{12},\label{64}\\
J^{(5)}_{13}&=J^{(1)}_{13},\label{65}\\
J^{(5)}_{21}&=\frac{-(1 + \theta_1) \left[R_1- T_1- (P_1 - S_1) \theta_1\right](R_0- T_0 -P_0 \theta_1 + S_0 \theta_1-\delta \rho_1)}{(1 - \delta)^2 \rho_1^2},\label{63}\\
J^{(5)}_{33}&=-J^{(1)}_{33}\label{66}.
\end{align}
The characteristic equation is given by
\begin{equation}
\left(J^{(5)}_{33}-\lambda\right)\left(\lambda^2-J^{(5)}_{11}\lambda-J^{(5)}_{12}J^{(5)}_{21}\right)=0\label{3char4}.
\end{equation}
Eigenvalue $\lambda_1 = J^{(5)}_{33} = \frac{\theta_1-\theta_{12}}{\theta_1+1}$ is negative if and only if $\theta_{12}>\theta_1$ (i.e., Eq.~\eqref{t12gtt1}). The real part of the other two eigenvalues is negative if and only if $-J^{(5)}_{11}>0$ and $-J^{(5)}_{12}J^{(5)}_{21}>0$. Recall that $- J^{(5)}_{11} = -J^{(1)}_{11}>0$ if and only if Eq. \eqref{stabineq} holds true.
Because $\rho_1$ and $\theta_1$ are positive, $J^{(5)}_{12}$ is negative.  Therefore, 
$-J^{(5)}_{12}J^{(5)}_{21}>0$ is equivalent to $J^{(5)}_{21} > 0$, which holds true if and only if $\delta < \delta_{\rm c,1}$ (i.e.,  Eq.~\eqref{deltan121}).

In sum, this face equilibrium is stable if and only if Eqs. \eqref{stabineq}, \eqref{t12gtt1}, and \eqref{deltan121} hold true.  

\subsection{Equilibrium $(x^\ast, n_1^\ast, n_{12}^\ast) = \left(\frac{1}{1+\theta_{12}},0,\frac{R_0-T_0-P_0\theta_{12}+S_0\theta_{12}}{\delta\rho_{12}}\right)$}

At $(x^\ast, n_1^\ast, n_{12}^\ast)=\left(\frac{1}{1+\theta_{12}},0,\frac{R_0-T_0-P_0\theta_{12}+S_0\theta_{12}}{\delta\rho_{12}}\right)$, the Jacobian is reduced to
\begin{equation}
\label{3J10}
J = \begin{pmatrix}
J^{(6)}_{11} &J^{(6)}_{12}  & J^{(6)}_{13}\\
0&J^{(6)}_{22}&0\\
J^{(6)}_{31}&0&0
\end{pmatrix},
\end{equation}
where
\begin{align}
J^{(6)}_{11}&=J^{(4)}_{11}\label{70},\\
J^{(6)}_{12}&=J^{(4)}_{12}\label{72},\\
J^{(6)}_{13}&=J^{(4)}_{13}\label{74},\\
J^{(6)}_{22}&=-J^{(4)}_{22}\label{73},\\
J^{(6)}_{31}&=\frac{-(1 + \theta_{12})\left[ R_0 -T_0 -(P_0 - S_0)\theta_{12} \right] (R_0 - T_0 - P_0\theta_{12} + S_0 \theta_{12}-\delta \rho_{12})}{\delta^2 \rho_{12}^2}\label{71}.
\end{align}
The characteristic equation is given by 
\begin{equation}
\left(J^{(6)}_{22}-\lambda\right)\left(\lambda^2-J^{(6)}_{11}\lambda-J^{(6)}_{13}J^{(6)}_{31}\right)=0\label{3char4}.
\end{equation}
Eigenvalue $\lambda_1 = J^{(6)}_{22} = \frac{\theta_1-\theta_{12}}{\theta_{12}+1}$ is negative if and only if $\theta_{12}>\theta_1$ (i.e., Eq.~\eqref{t12gtt1}). The real part of the other two eigenvalues is negative if and only if $-J^{(6)}_{11}>0$ and $-J^{(6)}_{13}J^{(6)}_{31}>0$. Recall that $- J^{(6)}_{11} = -J^{(4)}_{11}>0$ if and only if Eq. \eqref{stabineq} holds true.
Because $\rho_{12}$ and $\theta_{12}$ are positive, $J^{(6)}_{13}$ is negative. Therefore, $-J^{(6)}_{13}J^{(6)}_{31}>0$ is equivalent to $J^{(6)}_{31} > 0$, which holds true if and only if $\delta > \delta_{\rm c,2}$ (i.e., Eq.~\eqref{deltan10}).

In sum, this face equilibrium is stable if and only if Eqs.~\eqref{stabineq}, \eqref{t12gtt1}, and \eqref{deltan10} hold true.

\section{Corner equilibria of the three-dimensional system with $\theta_1= \theta_{12}$} \label{app:C2}

By evaluating Eq. \eqref{3Jac} at each corner equilibrium, we obtain the same Jacobians as those in Appendix~\ref{app:C1} with $\theta_1=\theta_{12}$.  The stability analysis of these corner equilibria is the same as that in Appendix~\ref{app:C1}, and we find that each Jacobian has at least one positive eigenvalue and one negative eigenvalue. Therefore, each of these corner equilibria is a saddle.

\section{Interior equilibria of the three-dimensional system with $\theta_1=\theta_{12}$} \label{app:I2}

In this section, we derive the stability requirements for the line of interior equilibria, $L$.
By setting $\theta_1=\theta_{12}$ and $x^\ast=\frac{1}{1+\theta_1}$, we obtain the $(2, 2)$ and $(3, 3)$ entries of
the Jacobian given by Eq. \eqref{3Jac} as follows:
\begin{equation}
-n_1(\theta_1x+x-1)+(1-n_1)(\theta_1x+x-1)
	=0
\end{equation}
and
\begin{equation}
-n_{12}(\theta_1x+x-1)+(1-n_{12})(\theta_1x+x-1) = 0.
\label{26}
\end{equation}
Therefore, we obtain
\begin{equation}\label{degjac2}
J(x,n_1,n_{12})=\begin{pmatrix}
x(1-x)\frac{\partial g}{\partial x} & x(1-x)\frac{\partial g}{\partial n_1} & x(1-x)\frac{\partial g}{\partial n_{12}}\\
+(1-2x)g(x,n_1,n_{12})&&\\
n_1(1-n_1)(1+\theta_1)&0&0\\
n_{12}(1-n_{12})(1+\theta_1)&0&0
\end{pmatrix}.
\end{equation}
By substituting $x^\ast=\frac{1}{1+\theta_1}$, we obtain
\begin{align}
\frac{\partial g}{\partial n_1}&=\frac{\partial q_{11}}{\partial n_1}-\frac{\partial q_{12}}{\partial n_1}\nonumber \\
&=-x(R_3-S_3-T_3+P_3)-(S_3-P_3)+\delta(R_3-T_3-S_3+P_3)x \nonumber\\ 
%
%
%
&=\frac{-(R_3-T_3)(1-\delta)-(S_3-P_3)(\theta_1+\delta)}{1+\theta_1}.
\label{28}
\end{align}
Therefore, we obtain
\begin{align}
x(1-x)\frac{\partial g}{\partial n_1}&=\frac{1}{1+\theta_1}\left(1-\frac{1}{1+\theta_1}\right)\frac{-(R_3-T_3-P_3\theta_1+S_3\theta_1)+\delta(R_3-S_3-T_3+P_3)}{1+\theta_1}\nonumber\\
%
%
%
&=\frac{-\theta_1[(R_3-T_3)(1-\delta)+(S_3-P_3)(\theta_1+\delta)]}{(\theta_1+1)^3} \notag\\
&\equiv \sigma_2\label{29}.
\end{align}
Likewise, using $x^\ast=\frac{1}{1+\theta_1}$, we obtain
\begin{align}
\frac{\partial g}{\partial n_{12}}&=\frac{-\delta[(R_3-T_3-S_3+P_3)+(S_3-P_3)(1+\theta_1)]}{1+\theta_1}\label{30}
\end{align}
and
\begin{align}
x(1-x)\frac{\partial g}{\partial n_{12}}&=\frac{1}{1+\theta_1}\left(1-\frac{1}{1+\theta_1}\right) \frac{-\delta(R_3-T_3+S_3\theta_1-P_3\theta_1)}{1+\theta_1} \nonumber\\
%
%
&=\frac{-\delta\theta_1(R_3-T_3+S_3\theta_1-P_3\theta_1)}{(\theta_1+1)^3} \notag\\
&\equiv \sigma_3 \label{31}.
\end{align}
Now, let us calculate the quantity for $x(1-x)\frac{\partial g}{\partial x}+(1-2x)g(x,n_1,n_{12})$. We obtain
\begin{align}
\frac{\partial g}{\partial x}&=\frac{\partial q_{11}}{\partial x}-\frac{\partial q_{12}}{\partial x}\nonumber\\
	&=(R_0-T_0-S_0+P_0)-(R_3-S_3-T_3+P_3) \left[n_1(1-\delta)+n_{12}\delta \right].
\label{32}
\end{align}	
By substituting Eq.~\eqref{degline} in Eq.~\eqref{32}, we obtain
\begin{align}
\frac{\partial g}{\partial x}&=(R_0-T_0-S_0+P_0)-(R_3-S_3-T_3+P_3) \frac{R_0-T_0-P_0\theta_1+S_0\theta_1}{R_3-T_3-P_3\theta_1+S_3\theta_1} \nonumber\\
	&=\frac{(\theta_1+1)[(P_1-S_1)(R_0-T_0)+(R_1-T_1)(S_0-P_0)]}{R_0-R_1-T_0+T_1-P_0\theta_1+P_1\theta_1+S_0\theta_1-S_1\theta_1}\label{33}.
\end{align}
Next, using $x^\ast=\frac{1}{1+\theta_1}$ and Eq. \eqref{degline}, we find
\begin{align}
	g(x,n_1,n_{12})&=q_{11}-q_{12}\nonumber\\
	&=(R_0-T_0-S_0+P_0)x+S_0-P_0 \nonumber\\
      &\;\;\;-\left[n_1(1-\delta)+\delta n_{12}\right][(R_3-S_3-T_3+P_3)x+(S_3-P_3)] \nonumber\\ 
&= 0.
\label{34}
\end{align}
Using Eqs. \eqref{33} and \eqref{34}, we obtain
\begin{align}
x(1-x)\frac{\partial g}{\partial x} =& \frac{\theta_1[(P_1-S_1)(R_0-T_0)+(R_1-T_1)(S_0-P_0)]}{(\theta_1+1)(R_0-R_1-T_0+T_1-P_0\theta_1+P_1\theta_1+S_0\theta_1-S_1\theta_1)} \notag\\
\equiv& \sigma_1. \label{35}
\end{align}
Using Eqs. \eqref{29}, \eqref{31}, and \eqref{35}, we find that the Jacobian at any point of $L$ is given by
\begin{equation}
J=\begin{pmatrix}
\sigma_1 & \sigma_2 & \sigma_3\\
\sigma_4&0&0\\
\sigma_5&0&0
\end{pmatrix},
\label{degjac3}
\end{equation}
where 
\begin{align}
\sigma_4&=n_1^\ast(1-n_1^\ast)(1+\theta_1)\label{39},\\
\sigma_5&=n_{12}^\ast(1-n_{12}^\ast)(1+\theta_1).\label{40}
\end{align}

The characteristic equation is given by
\begin{equation}
\lambda\left[\lambda^2-\sigma_1\lambda-(\sigma_2 \sigma_4+\sigma_3 \sigma_5)\right]=0\label{3charpoly}.
\end{equation}
Eigenvalue $\lambda_1=0$ reflects the fact that the line of equilibria, $L$, is neutrally stable along the direction of $L$. The other two eigenvalues, $\lambda_2$ and $\lambda_3$, are given by 
\begin{equation}\label{3eig}
\lambda_{2,3}=\frac{\sigma_1\pm \sqrt{\sigma_1^2+4(\sigma_2 \sigma_4+\sigma_3 \sigma_5)}}{2}.
\end{equation}
Let $\alpha=-\sigma_1$ and $\beta=-\sigma_2 \sigma_4-\sigma_3 \sigma_5$. The real part of $\lambda_2$ and $\lambda_3$ is negative if and only if $\alpha>0$ and $\beta>0$.  Because the denominator of Eq. \eqref{35} is positive, then $\alpha>0$ if and only if Eq. \eqref{stabineq} holds true.

Now we seek the conditions under which $\beta>0$. Because $n_1^\ast$ and $n_{12}^\ast$ are positive, we obtain
$\sigma_4 > 0$ and $\sigma_5 > 0$.
%
%
Therefore, a sufficient condition for $\beta > 0$ is that both $\sigma_2$ and $\sigma_3$ are negative. Using the assumptions in Eq. \eqref{inequal}, we find that the numerators of $\sigma_2$ and $\sigma_3$ are always negative.  Thus, under Eqs.~\eqref{inequal} and \eqref{stabineq}, we obtain $\alpha>0$ and $\beta>0$ such that the real parts of $\lambda_2$ and $\lambda_3$ are negative.

\section{Edge equilibria of the three-dimensional system with $\theta_1 =\theta_{12}$} \label{app:E2}

\subsection{Equilibrium $(x^\ast, n_1^\ast, n_{12}^\ast)=\left(\frac{P_0-S_0-\delta(P_0-P_1-S_0+S_1)}{R_0-T_0-S_0+P_0-\delta\gamma},0,1\right)$}\label{app:E22}

For the equilibrium $(x^\ast, n_1^\ast, n_{12}^\ast)=\left(\frac{P_0-S_0-\delta(P_0-P_1-S_0+S_1)}{R_0-T_0-S_0+P_0-\delta\gamma},0,1\right)$, the Jacobian, Eq.~\eqref{3Jac}, is reduced to
\begin{equation}
J = \begin{pmatrix}
J^{(7)}_{11} &J^{(7)}_{12}  & J^{(7)}_{13}\\
0&J^{(7)}_{22}&0\\
0&0&J^{(7)}_{33}
\end{pmatrix},
\end{equation}
where
\begin{align}
J^{(7)}_{11}&=\frac{[(P_0-S_0)(1-\delta)+\delta(P_1-S_1)][(R_0-T_0)(1-\delta)+\delta(R_1-T_1)]}{R_0-T_0-S_0+P_0-\delta\gamma}\label{86},\\
J^{(7)}_{12}&=\frac{(1-\delta)\omega[(P_0-S_0)(1-\delta)+\delta(P_1-S_1)][(R_0-T_0)(1-\delta)+\delta(R_1-T_1)]}{(R_0-T_0-S_0+P_0-\delta\gamma)^3}\label{87},\\
J^{(7)}_{13}&=\frac{\delta\omega[(P_0-S_0)(1-\delta)+\delta(P_1-S_1)][(R_0-T_0)(1-\delta)+\delta(R_1-T_1)]}{(R_0-T_0-S_0+P_0-\delta\gamma)^3}\label{88},\\
J^{(7)}_{22}&=-\frac{R_0-T_0-P_0\theta_1+S_0\theta_1-\delta\rho_1}{R_0-T_0-S_0+P_0-\delta\gamma}\label{89},\\
J^{(7)}_{33}&=-J^{(7)}_{22},\label{90}
\end{align}
and
\begin{equation}
\omega= (P_0-S_0)(R_1-T_1)-(P_1-S_1)(R_0-T_0)
\label{omeg}.
\end{equation}

The eigenvalues of $J$ are given by $\lambda = J^{(7)}_{11}, J^{(7)}_{22}$, and $J^{(7)}_{33}$.
Using the fact that each eigenvalue has a common denominator, which we refer to as
\begin{equation}
\mu_3 \equiv R_0-T_0-S_0+P_0-\delta\gamma,
\end{equation}
we determine the stability of the edge equilibrium by examining the following four possible cases.

\subsubsection*{Case 1: $\gamma>0$ and $\mu_3>0$}

First, $\gamma$ is positive if and only if 
\begin{equation}\label{gampos}
R_0-T_0-R_1+T_1>S_0-P_0-S_1+P_1.
\end{equation}
Under this condition, $\mu_3$ is positive if and only if
\begin{equation}\label{m3pos}
\delta<\frac{R_0-T_0-S_0+P_0}{\gamma}.
\end{equation}
For positive $\delta$ values satisfying Eq.~\eqref{m3pos} to exist, it must hold true that $R_0-T_0>S_0-P_0$. 

Eigenvalues $J^{(7)}_{22}$ and $J^{(7)}_{33}$ are non-positive if and only if 
\begin{align}
\delta&\leq\frac{R_0-T_0-P_0\theta_1+S_0\theta_1}{\rho_1}\label{200}
\intertext{and}
\delta&\geq\frac{R_0-T_0-P_0\theta_1+S_0\theta_1}{\rho_1}\label{201},
\end{align}
respectively, which implies that 
\begin{equation}
\delta=\frac{R_0-T_0-P_0\theta_1+S_0\theta_1}{\rho_1}\label{delcon1}.
\end{equation} 
Lastly, eigenvalue $J^{(7)}_{11}$ is negative if either
\begin{align}
\delta< \min \left\{ \frac{P_0-S_0}{P_0-S_0-P_1+S_1}, \frac{R_0-T_0}{R_0-T_0-R_1+T_1} \right\}\label{202}
\end{align}
or
\begin{align}
\delta > \max \left\{ \frac{P_0-S_0}{P_0-S_0-P_1+S_1}, \frac{R_0-T_0}{R_0-T_0-R_1+T_1} \right\}\label{203}
\end{align}
holds true. However, we find that there is no $\delta$ value that simultaneously satisfies Eqs.~\eqref{m3pos}, \eqref{delcon1}, and \eqref{202}, or one that simultaneously satisfies Eqs.~\eqref{m3pos}, \eqref{delcon1}, and \eqref{203}.

\subsubsection*{Case 2: $\gamma>0$ and $\mu_3 < 0$}

If $\gamma > 0$, then $\mu_3 < 0$ if and only if 
\begin{equation}\label{m3neg}
\delta>\frac{R_0-T_0-S_0+P_0}{\gamma}.
\end{equation}
For positive $\delta$ values satisfying Eq.~\eqref{m3neg} to exist and be less than $1$, it must hold true that $R_1-T_1<S_1-P_1$. Eigenvalues $J^{(7)}_{22}$ and $J^{(7)}_{33}$ are non-positive if and only if
\begin{align}
\delta\geq&\frac{R_0-T_0-P_0\theta_1+S_0\theta_1}{\rho_1} \label{204}
\intertext{and}
\delta\leq&\frac{R_0-T_0-P_0\theta_1+S_0\theta_1}{\rho_1}, \label{205}
\end{align}
respectively,  which implies that 
\begin{equation}
\delta=\frac{R_0-T_0-P_0\theta_1+S_0\theta_1}{\rho_1}\label{delcon2}.
\end{equation} 
Because we have assumed that $\mu_3< 0$, the numerator of Eq.~\eqref{86} has to be positive for eigenvalue $J^{(7)}_{11}$ to be negative. Then, either both $(P_0-S_0)(1-\delta)+\delta(P_1-S_1)$ and $(R_0-T_0)(1-\delta)+\delta(R_1-T_1)$ are positive or both are negative.  

If both $(P_0-S_0)(1-\delta)+\delta(P_1-S_1)$ and $(R_0-T_0)(1-\delta)+\delta(R_1-T_1)$ are positive, we obtain
\begin{align}
\delta&>\frac{P_0-S_0}{P_0-S_0-P_1+S_1}\label{206}
\intertext{and}
\delta&<\frac{R_0-T_0}{R_0-T_0-R_1+T_1}.\label{207}
\end{align}
We find that there is no $\delta$ value that simultaneously satisfies Eqs.~\eqref{m3neg}, \eqref{delcon2}, \eqref{206}, and \eqref{207}.

If both $(P_0-S_0)(1-\delta)+\delta(P_1-S_1)$ and $(R_0-T_0)(1-\delta)+\delta(R_1-T_1)$ are negative, we obtain
\begin{align}
\delta&<\frac{P_0-S_0}{P_0-S_0-P_1+S_1}\label{208}
\intertext{and}
\delta&>\frac{R_0-T_0}{R_0-T_0-R_1+T_1}.\label{209}
\end{align}
A $\delta$ value satisfying Eqs.~\eqref{208} and \eqref{209} exists if and only if
\begin{equation}
\frac{R_0-T_0}{R_0-T_0-R_1+T_1}<\frac{P_0-S_0}{P_0-S_0-P_1+S_1},
\end{equation}
which is equivalent to Eq.~\eqref{stabineq}.

Such a $\delta$ value satisfying $\delta \in (0, 1)$ exists if and only if the RHS of Eq.~\eqref{208} is positive and that of Eq.~\eqref{209} is less than $1$.
In fact, because $P_0-S_0<0$ and $S_1-P_1<0$, the numerator and denominator on the RHS of Eq.~\eqref{208} are negative, which implies that the RHS is positive. The RHS of Eq.~\eqref{209} is less than $1$ because
we obtain $R_0-T_0 > 0$ and $R_1 - T_1 < 0$ from Eq.~\eqref{inequal}. Lastly, the intersection of the conditions derived for $\delta$, given by Eqs.~\eqref{m3neg}, \eqref{delcon2}, \eqref{208}, and \eqref{209}, yields Eq.~\eqref{0,1stabcon}, i.e., $\delta_{\rm c,1} = \delta_{\rm c,2}$.

\subsubsection*{Case 3: $\gamma< 0$ and $\mu_3 > 0$}

In this section, we assume that $\gamma < 0$, i.e
\begin{equation}\label{gamneg}
R_0-T_0-R_1+T_1<S_0-P_0-S_1+P_1.
\end{equation}
Then, $\mu_3 > 0$ if and only if Eq.~\eqref{m3neg} is satisfied. For positive $\delta$ values satisfying Eq.~\eqref{m3neg} to exist and be less than $1$, it must hold true that $R_1-T_1<S_1-P_1$. If $\mu_3 > 0$, then we obtain Eqs.~\eqref{delcon1}, \eqref{202}, and \eqref{203}, and again find that there is no $\delta$ value that simultaneously satisfies these inequalities.

\subsubsection*{Case 4: $\gamma<0$ and $\mu_3 < 0$}

In this section, we assume that $\gamma < 0$ and $\mu_3< 0$, which requires Eq.~\eqref{m3pos} to be satisfied. For positive $\delta$ values satisfying Eq.~\eqref{m3pos} to exist, it must hold true that $R_0-T_0>S_0-P_0$. The derivation of the stability of this case follows the same derivation as Case 2, and we find that the equilibrium is marginally stable if and only if Eqs. \eqref{stabineq} and \eqref{0,1stabcon} hold true.

\subsection{Equilibrium $(x^\ast, n_1^\ast, n_{12}^\ast)=\left(\frac{P_1-S_1+\delta(P_0-P_1-S_0+S_1)}{R_1-T_1-S_1+P_1+\delta\gamma},1,0\right)$}\label{app:E22}

For the equilibrium $(x^\ast, n_1^\ast, n_{12}^\ast)=\left(\frac{P_1-S_1+\delta(P_0-P_1-S_0+S_1)}{R_1-T_1-S_1+P_1+\delta\gamma},1,0\right)$, the Jacobian, Eq.~\eqref{3Jac}, is reduced to
\begin{equation}
J = \begin{pmatrix}
J^{(8)}_{11} &J^{(8)}_{12}  & J^{(8)}_{13}\\
0&J^{(8)}_{22}&0\\
0&0&J^{(8)}_{33}
\end{pmatrix},
\end{equation}
where
\begin{align}
J^{(8)}_{11}&=\frac{[(P_1-S_1)(1-\delta)+\delta(P_0-S_0)][(R_1-T_1)(1-\delta)+\delta(R_0-T_0)]}{R_1-T_1-S_1+P_1+\delta\gamma}\label{91},\\
J^{(8)}_{12}&=\frac{(1-\delta)\omega[(P_1-S_1)(1-\delta)+\delta(P_0-S_0)][(R_1-T_1)(1-\delta)+\delta(R_0-T_0)]}{(R_1-T_1-S_1+P_1+\delta\gamma)^3}\label{92},\\
J^{(8)}_{13}&=\frac{\delta\omega[(P_1-S_1)(1-\delta)+\delta(P_0-S_0)][(R_1-T_1)(1-\delta)+\delta(R_0-T_0)]}{(R_0-T_0-S_0+P_0-\delta\gamma)^3}\label{93},\\
J^{(8)}_{22}&=-\frac{R_1-T_1-P_1\theta_1+S_1\theta_1+\delta\rho_1}{R_1-T_1-S_1+P_1+\delta\gamma}\label{94},\\
J^{(8)}_{33}&=-J^{(8)}_{22}.\label{95}
\end{align}
The eigenvalues of $J$ are given by $\lambda = J^{(8)}_{11}$, $J^{(8)}_{22}$, and $J^{(8)}_{33}$. Similarly to the prior equilibrium, each eigenvalue has a common denominator, which we refer to as
\begin{equation}
\mu_4 \equiv R_1-T_1-S_1+P_1+\delta\gamma.
\end{equation}
We determine the stability of the edge equilibrium by examining the following four possible cases.

\subsubsection*{Case 1: $\gamma>0$ and $\mu_4>0$} 

First, $\gamma$ is positive if and only if Eq.~\eqref{gampos} is satisfied. When $\gamma > 0$ is satisfied, $\mu_4$ is positive if and only if
\begin{equation}\label{m4pos}
\delta>\frac{-(R_1-T_1-S_1+P_1)}{\gamma}.
\end{equation}
For positive $\delta$ values satisfying Eq.~\eqref{m4pos} to exist and be less than $1$, it must hold true that $R_0-T_0>S_0-P_0$. 

Eigenvalues $J^{(8)}_{22}$ and $J^{(8)}_{33}$ are non-positive if and only if 
\begin{align}
\delta&\leq\frac{-(R_1-T_1-P_1\theta_1+S_1\theta_1)}{\rho_1}
\label{210}
\intertext{and}
\delta&\geq\frac{-(R_1-T_1-P_1\theta_1+S_1\theta_1)}{\rho_1},\label{211}
\end{align}
respectively, which implies that 
\begin{equation}
\delta=\frac{-(R_1-T_1-P_1\theta_1+S_1\theta_1)}{\rho_1}\label{delcon3}.
\end{equation}   Lastly, $J^{(8)}_{11}$ is negative if either
\begin{align}
\delta< \min \left\{ \frac{-(P_1-S_1)}{P_0-S_0-P_1+S_1},\frac{-(R_1-T_1)}{R_0-T_0-R_1+T_1}\right \}\label{212}
\end{align}
or
\begin{align}
\delta > \max \left\{\frac{-(P_1-S_1)}{P_0-S_0-P_1+S_1},\frac{-(R_1-T_1)}{R_0-T_0-R_1+T_1}\right \}\label{213}
\end{align}
holds true.  However, we find that there is no $\delta$ value that simultaneously satisfies Eqs.~\eqref{m4pos}, \eqref{delcon3}, and \eqref{212} or Eqs.~\eqref{m4pos}, \eqref{delcon3}, and \eqref{213}.

\subsubsection*{Case 2: $\gamma >0$ and $\mu_4<0$} 

If $\gamma$ is positive, then $\mu_4<0$ if and only if 
\begin{equation}\label{m4neg}
\delta<\frac{-(R_1-T_1-S_1+P_1)}{\gamma}.
\end{equation}
For positive $\delta$ values satisfying Eq.~\eqref{m4neg} to exist, it must hold true that $R_1-T_1<S_1-P_1$. Eigenvalues $J^{(8)}_{22}$ and $J^{(8)}_{33}$ are non-positive if and only if
\begin{align}
\delta\geq&\frac{-R_1+T_1+P_1\theta_1-S_1\theta_1}{\rho_1} \label{214}
\intertext{and}
\delta\leq&\frac{-R_1+T_1+P_1\theta_1-S_1\theta_1}{\rho_1}, \label{215}
\end{align}
respectively, which implies that 
\begin{equation}
\delta=\frac{-(R_1-T_1-P_1\theta_1+S_1\theta_1)}{\rho_1}\label{delcon4}.
\end{equation} 
Because we have assumed that $\mu_4<0$, the numerator of Eq.~\eqref{91} has to be positive for eigenvalue $J^{(8)}_{11}$ to be negative.  Then, either both $(P_1-S_1)(1-\delta)+\delta(P_0-S_0)$ and $(R_1-T_1)(1-\delta)+\delta(R_0-T_0)$ are positive or both are negative.  

If both $(P_1-S_1)(1-\delta)+\delta(P_0-S_0)$ and $(R_1-T_1)(1-\delta)+\delta(R_0-T_0)$ are positive, we obtain
\begin{align}
\delta&<\frac{-(P_1-S_1)}{P_0-S_0-P_1+S_1}\label{216}
\intertext{and}
\delta&>\frac{-(R_1-T_1)}{R_0-T_0-R_1+T_1}\label{217}.
\end{align}
We find that there is no $\delta$ value that satisfies Eqs.~\eqref{m4neg}, \eqref{delcon4}, \eqref{216}, and \eqref{217}.

If both $(P_1-S_1)(1-\delta)+\delta(P_0-S_0)$ and $(R_1-T_1)(1-\delta)+\delta(R_0-T_0)$ are negative, we obtain
\begin{equation}\label{218}
\delta>\frac{-(P_1-S_1)}{P_0-S_0-P_1+S_1}
\end{equation}
and 
\begin{equation}\label{219}
\delta<\frac{-(R_1-T_1)}{R_0-T_0-R_1+T_1}.
\end{equation}
A value of $\delta$ satisfying Eqs.~\eqref{218} and \eqref{219} exists if and only if
\begin{equation}
\frac{-(R_1-T_1)}{R_0-T_0-R_1+T_1}<\frac{-(P_1-S_1)}{P_0-S_0-P_1+S_1},
\end{equation}
which is equivalent to Eq. \eqref{stabineq}.  Such a $\delta$ value satisfying $\delta \in (0, 1)$ exists if and only if 
the RHS of Eq.~\eqref{218} is less than $1$ and that of Eq.~\eqref{219} is positive.
In fact, the RHS of Eq.~\eqref{218} is less than $1$ because we obtain $P_0-S_0 < 0$ and $P_1 - S_1 > 0$ from Eq.~\eqref{inequal}. Because $R_0-T_0>0$ and $T_1-R_1>0$, the numerator and denominator on the RHS of Eq.~\eqref{219} are positive, which implies that the RHS is positive. Lastly, the intersection of the conditions derived for $\delta$, given by Eqs.~\eqref{m4neg}, \eqref{delcon4}, \eqref{218}, and \eqref{219}, yields Eq.~\eqref{1,0stabcon}, i.e., $\delta_{\rm c,3} = \delta_{\rm c,4}$.

\subsubsection*{Case 3: $\gamma<0$ and $\mu_4>0$} 

In this section, we assume that $\gamma<0$, i.e., Eq.~\eqref{gamneg}. Then, $\mu_4>0$ if and only if Eq.~\eqref{m4neg} is satisfied. For positive $\delta$ values satisfying Eq.~\eqref{m4neg} to exist, it must hold true that $R_1-T_1<S_1-P_1$. If $\mu_4>0$, then we obtain Eqs.~\eqref{delcon3}, \eqref{212}, \eqref{213}, and \eqref{m4neg}, and find that there is no $\delta$ value that simultaneously satisfies these inequalities.

\subsubsection*{Case 4: $\gamma<0$ and $\mu_4<0$} 

In this section, we assume that $\gamma<0$ and $\mu_4<0$, which requires Eq.~\eqref{m4pos}. For positive $\delta$ values satisfying Eq.~\eqref{m4pos} to exist and be less than $1$, it must hold true that $R_0-T_0>S_0-P_0$. The derivation of the stability of this case follows the same derivation as Case 2, and we find that the equilibrium is marginally stable if and only if Eqs.~\eqref{stabineq} and \eqref{1,0stabcon} hold true.

\section{Equilibria of the five-dimensional system} \label{app:ST}

We show the $60$ equilibria of the five-dimensional system with their stability requirements in Table \ref{5dimtable}. 

\begin{table}[!h]
\center
\caption{Equilbria of the five-dimensional system with their stability requirements. Symbol ``a'' represents an equilibrium value between $0$ and $1$. When the analytical expression is too complicated, we show the numerical values to the third significant digit.}
\def\arraystretch{1.7}
\begin{tabular}{|c|c|c|c|c||c|}
\hline
$x^\ast$ & $y^\ast$ & $n_1^\ast$ & $n_2^\ast$ & $n_{12}^\ast$&Stability conditions \\ \hline\hline
        0 & a & 0 & 0 & 1&$S_0-P_0>R_0-T_0, R_1-T_1<S_1-P_1, \theta_{12}>1+2\theta_2, \frac{1+2\theta_2}{3+4\theta_2}<\delta<\frac{\theta_{12}}{1+2\theta_{12}}$
 \\ 
\hline
        0 & a & 0 & 0 & a&$S_0-P_0>R_0-T_0, R_1-T_1<S_1-P_1, \delta>\frac{\theta_{12}}{1 + 2\theta_{12}}, \theta_{12}>1+2\theta_2$ \\ 
\hline
        0 & a & 0 & 1 & 0 & Never stable\\ 
\hline
        0 & a & 0 & 1 & a&$S_0-P_0>R_0-T_0, R_1-T_1<S_1-P_1, 1<\theta_{12}< 1 + 2 \theta_2,  \delta>\frac{91+52\theta_{12}+9\theta_{12}^2}{93+54\theta_{12}+9\theta_{12}^2}$ \\ 
        ~ & ~ & ~ & ~ & ~ &$S_0-P_0<R_0-T_0, R_1-T_1<S_1-P_1, 1<\theta_{12}<1+2\theta_2, \delta>\frac{91+52\theta_{12}+9\theta_{12}^2}{93+54\theta_{12}+9\theta_{12}^2}$\\ 
\hline
        0 & a & 0 & a & 0 &Never stable\\ \hline
        0 & a & 0 & a & 1 & $S_0-P_0>R_0-T_0, R_1-T_1<S_1-P_1, \theta_2<\frac{1}{2}(-1+\theta_{12}), 0.471<\delta<\frac{1+2\theta_2}{3+4\theta_2}$\\ 
\hline
        0 & a & 1 & 0 & 1 &Never stable\\ 
\hline
        0 & a & 1 & 0 & a&Never stable\\ \hline
        0 & a & 1 & 1 & 0 &Never stable \\ 
\hline
        0 & a & 1 & 1 & a&Never stable \\ \hline
        0 & a & 1 & a & 0 &Never stable\\ \hline
        0 & a & 1 & a & 1 &Never stable\\ \hline
        1 & a & 0 & 0 & 1&Never stable \\ 
\hline
        1 & a & 0 & 0 & a &Never stable\\ \hline
        1 & a & 0 & 1 & 0&Never stable \\ 
\hline
        1 & a & 0 & 1 & a&Never stable \\ \hline
        1 & a & 0 & a & 0&Never stable \\ \hline
\end{tabular}
\label{5dimtable}
\end{table}
\begin{table}
\center
\def\arraystretch{1.7}
\begin{tabular}[!h]{|c|c|c|c|c||c|}
\hline
       1 & a & 0 & a & 1&Never stable \\ \hline
        1 & a & 1 & 0 & 1&Never stable \\ \hline
 1 & a & 1 & 0 & a &$S_0-P_0>R_0-T_0, R_1-T_1<S_1-P_1, \frac{\theta_2}{2 + \theta_2}<\theta_{12}<1,\delta>\frac{9+20\theta_{12}+9\theta_{12}^2}{9+24\theta_{12}+13\theta_{12}^2}$\\ 
\hline
        1 & a & 1 & 1 & 0 &$S_0-P_0>R_0-T_0, R_1-T_1<S_1-P_1, \theta_2>2, \theta_{12}<\frac{1}{2}, \frac{3}{4}<\delta<\frac{5-\theta_{12}}{6}$\\ \hline
        1 & a & 1 & 1 & a & $S_0-P_0<R_0-T_0, R_1-T_1<S_1-P_1, \theta_{12}<\frac{\theta_2}{2+\theta_2},\delta>\frac{41+17\theta_{12}}{42+18\theta_{12}}$ \\ 
        ~ & ~ & ~ & ~ & ~& $S_0-P_0>R_0-T_0, R_1-T_1<S_1-P_1, \theta_2>2, \theta_{12}<\frac{1}{2}, \frac{5-\theta_{12}}{6}<\delta<0.766$\\ 
\hline
        1 & a & 1 & a & 0&$S_0-P_0>R_0-T_0, R_1-T_1<S_1-P_1, \theta_2<2, \frac{5(1+\theta_2)}{6 + 7 \theta_2}<\delta<\frac{5+2\theta_2}{6+3\theta_2}, \theta_{12}<0.454$\\ 
\hline
        1 & a & 1 & a & 1 &Never stable\\ \hline
        a & 0 & 0 & 0 & 1 &$S_0-P_0 > R_0-T_0, R_1-T_1 < S_1-P_1, \theta_{12}>1, 1+2\theta_1<\theta_{12}, \frac{1+2\theta_1}{3+4\theta_1}<\delta<\frac{\theta_{12}}{1+2\theta_{12}}$\\ 
\hline
        a& 0 & 0 & 0 & a&$S_0-P_0>R_0-T_0, R_1-T_1<S_1-P_1, \theta_{12}>1+2\theta_1, \delta>\frac{\theta_{12}}{1+2\theta_{12}}$\\ 
\hline
        a & 0 & 0 & 1 & 1&Never stable\\ 
\hline
         a & 0 & 0 & 1 & a &Never stable\\ \hline
        a & 0 & 1 & 0 & 0 &Never stable\\ 
\hline
          a & 0 & 1 & 0 & a& $S_0-P_0<R_0-T_0, R_1-T_1<S_1-P_1, 1<\theta_{12}<1+2\theta_1,\delta>\frac{91+52\theta_{12}+9\theta_{12}^2}{93+54\theta_{12}+9\theta_{12}^2}$ \\
        ~ & ~ & ~ & ~ & ~ & $S_0-P_0>R_0-T_0, R_1-T_1< S_1-P_1, 1<\theta_{12}<1+2\theta_1, \delta>\frac{9+8\theta_{12}+2\theta_{12}^2}{11+10\theta_{12}+2\theta_{12}^2}$\\ 
\hline
         a & 0 & 1 & 1 & 0&Never stable\\ 
\hline
        a & 0 & 1 & 1 & a &Never stable\\ \hline
         a & 0 & a & 0 & 0&Never stable \\ \hline
        a & 0 & a & 0 & 1 &$S_0-P_0>R_0-T_0, R_1-T_1<S_1-P_1, \theta_{12}>1, \theta_{12}>1+2\theta_1, 0.43<d<\frac{1+2\theta_1}{3+4\theta_1}$\\ \hline
        a & 0 & a & 1 & 0 &Never stable\\ \hline
        a & 0 & a & 1 & 1 &Never stable\\ \hline
        a & 1 & 0 & 0 & 1 &Never stable\\ \hline
        a & 1 & 0 & 0 & a &Never stable\\ \hline
        a & 1 & 0 & 1 & 1 &Never stable\\ \hline
        a & 1 & 0 & 1 & a &$S_0-P_0>R_0-T_0, R_1-T_1<S_1-P_1, \frac{\theta_1}{2+\theta_1}<\theta_{12}<1, \delta>\frac{3+20\theta_{12}+9\theta_{12}^2}{3+24\theta_{12}+13\theta_{12}^2}$\\ \hline
\end{tabular}
\end{table}
\begin{table}
\center
\def\arraystretch{1.7}
\hspace*{-1cm}
\begin{tabular}[!h]{|c|c|c|c|c||c|}
\hline
a & 1 & 1 & 0 & 0 &Never stable\\ \hline
         a & 1 & 1 & 0 & a &Never stable\\ \hline
        a & 1 & 1 & 1 & 0 &$S_0-P_0>R_0-T_0, R_1-T_1<S_1-P_1, \theta_{12}<\frac{1}{2}, \theta_1>2, \frac{3}{4}<\delta<\frac{5-\theta_{12}}{6}$\\ \hline 
       a & 1 & 1 & 1 & a &$S_0-P_0<R_0-T_0, R_1-T_1<S_1-P_1, 0<\theta_{12}<\frac{\theta_1}{2+\theta_1}, \delta>\frac{41+17\theta_{12}}{42+18\theta_{12}}$\\ 
        ~ & ~ & ~ & ~ & ~& $S_0-P_0>R_0-T_0, R_1-T_1<S_1-P_1, \theta_1>2.71, \theta_{12}<0.574, \delta>0.76667$\\ \hline
        a & 1 & a & 0 & 0 &Never stable\\ \hline
        a & 1 & a & 0 & 1&Never stable \\ \hline
        a & 1 & a & 1 & 0 &$S_0-P_0>R_0-T_0, R_1-T_1<S_1-P_1,\theta_{12}<0.454, 1.66<\theta_1<2, \frac{5+5\theta_1}{6+7\theta_1}<\delta<\frac{5+2\theta_1}{6+3\theta_1}$\\ \hline
        a & 1 & a & 1 & 1&Never stable \\ \hline
        a & a & 0 & 0 & 1&$S_0-P_0>R_0-T_0, R_1-T_1<S_1-P_1, \theta_2<\frac{1}{2}, \theta_2<\theta_{12}<\frac{1}{2}, \theta_1<\theta_2, \frac{1+2\theta_2}{6+4\theta_2}<\delta<\frac{1+2\theta_{12}}{6+4\theta_{12}}$ \\ \hline
        a & a & 0 & a & 1 &Never stable\\ \hline
        a & a & 0 & a & a &Never stable\\ \hline
        a & a & 1 & 1 & 0 &$S_0-P_0<R_0-T_0, R_1-T_1<S_1-P_1, \theta_{12}<\theta_2<\theta_1, \frac{5+2\theta_2}{7+3\theta_2}<\delta<\frac{5+2\theta_{12}}{7+3\theta_{12}}$ \\ 
        ~ & ~ & ~ & ~ & ~&$S_0-P_0>R_0-T_0, R_1-T_1<S_1-P_1, \theta_2>\frac{1}{2}, \theta_{12}<\frac{1}{2}, \theta_1>\theta_2, \frac{5+2\theta_2}{6+4\theta_2}<\delta<\frac{3}{4}$ \\ \hline
        a & a & 1 & a & 0 &$S_0-P_0<R_0-T_0, R_1-T_1<S_1-P_1, \theta_{12}<\theta_2<\theta_1, \frac{5+2\theta_{12}}{7+3\theta_{12}}<\delta<\frac{1+2\theta_1}{1+3\theta_1}$\\ 
        ~ & ~ & ~ & ~ & ~ &$S_0-P_0>R_0-T_0, R_1-T_1<S_1-P_1, \theta_2>2, \theta_{12}<\frac{1}{2}, 0.58<\delta<\frac{5+2\theta_2}{7+3\theta_2}$\\ \hline
        a & a & 1 & a & a &Never stable\\ \hline
        a & a & a & 0 & a&Never stable \\ \hline
        a & a & a & 0 & a &Never stable\\ \hline
        a & a & a & 1 & 0 &Never stable\\ \hline
        a & a & a & 1 & a&Never stable \\ \hline
        a & a & a & a & 0&$S_0-P_0<R_0-T_0, R_1-T_1<S_1-P_1, \theta_1+\theta_2+2\theta_1\theta_2>\theta_{12}(2+\theta_1+\theta_2), \delta<0.666$ \\ 
        ~ & ~ & ~ & ~ & ~&$S_0-P_0>R_0-T_0, R_1-T_1<S_1-P_1, \theta_1+\theta_2+2\theta_1\theta_2>\theta_{12}(2+\theta_1+\theta_2), \delta<0.58$\\ \hline
        a & a & a & a & 1&$S_0-P_0<R_0-T_0, R_1-T_1<S_1-P_1, \theta_1+\theta_2+2\theta_1\theta_2>\theta_{12}(2+\theta_1+\theta_2), \delta<0.04$ \\ 
         ~ & ~ & ~ & ~ & ~ &$S_0-P_0>R_0-T_0, R_1-T_1<S_1-P_1,\theta_1+\theta_2+2\theta_1\theta_2>\theta_{12}(2+\theta_1+\theta_2), \delta<0.288$\\ \hline
\end{tabular}
\end{table}
\end{appendices}
\end{document}